%% file: main.tex
\documentclass[letterpaper,twocolumn,10pt]{article}
\usepackage{usenix2019_v3}

\usepackage{tikz}
\usepackage{amsmath}
\usepackage{xspace}

\usepackage{balance}
\usepackage[normalem]{ulem}
\usepackage{graphicx}
\usepackage{algorithm}
\usepackage[noend]{algpseudocode}
\usepackage{amsthm}
\usepackage{amsmath}
\usepackage{multirow}
\usepackage{url}
\usepackage{subcaption}
\usepackage{enumitem}
\usepackage{makecell}
\usepackage{booktabs} 
\usepackage{soul}
\usepackage{textcomp}
\usepackage{pifont}
\usepackage{adjustbox}
\usepackage{wrapfig}
\usepackage[T1]{fontenc}
\usepackage{inconsolata}

\usepackage{amssymb}
\usepackage{bbm}
\usepackage{tabularx}
\usepackage{bbding}

\usepackage[normalem]{ulem}

\usepackage[most]{tcolorbox}

\usepackage{tikz}

\newcommand\wcircle[1]{\raisebox{-0.5ex}{\Large\ding{\numexpr171+#1}}}

\theoremstyle{definition}

\newcommand{\SystemName}{\texttt{RollPacker}\xspace}

\begin{document}

\date{}

% make title bold and 14 pt font (Latex default is non-bold, 16 pt)
% \title{\Large \bf \SystemName: Packing Long-Tail Rollouts for RL Post-Training with Tail Batching}

 \title{\Large \bf \SystemName: Mitigating Long-Tail Rollouts for Fast, Synchronous RL Post-Training}
 
%for single author (just remove % characters)
\author{
    {\rm Wei Gao$^{\dagger}$, Yuheng Zhao$^{\dagger}$, Dakai An$^{\dagger}$, Tianyuan Wu$^{\dagger}$, Lunxi Cao$^{\dagger}$,} \\
    {\rm Shaopan Xiong$^{\ddagger}$, Ju Huang$^{\ddagger}$, Weixun Wang$^{\S}$, Siran Yang$^{\ddagger}$, Wenbo Su$^{\S}$,} \\
    {\rm Jiamang Wang$^{\ddagger}$, Lin Qu$^{\ddagger}$, Bo Zheng$^{\ddagger}$, Wei Wang$^{\dagger}$}
    \bigskip \\
    {\em $^{\dagger}$Hong Kong University of Science and Technology} \\
    {\em $^{\ddagger}$Alibaba Group} \\
    {\em $^{\S}$Taobao \& Tmall Group of Alibaba}
     \vspace{0.5cm}
}
% Equal contribution footnote

\maketitle
\footnotetext[1]{Wei Gao and Yuheng Zhao contributed equally to this work.}

\maketitle

\input{contents/0abstract}

\input{contents/1introduction}

\input{contents/2background}
\input{contents/3designprinciples}
\input{contents/4systemdesign}

\input{contents/5implementation}
\input{contents/6evaluation}

\input{contents/7relatedworks}

\input{contents/8conclustion}

% \clearpage

\bibliographystyle{plain}
\bibliography{reference}
\end{document}

%% file: contents/0abstract.tex
%!TEX root=../main.tex
\begin{abstract}

Reinforcement Learning (RL) is a pivotal post-training technique for enhancing the reasoning capabilities of Large Language Models (LLMs). However, synchronous RL post‑training often suffers from significant GPU underutilization, referred to as “bubbles”, caused by imbalanced response lengths within rollout steps. Many RL systems attempt to alleviate this problem by relaxing synchronization, but this can compromise training accuracy. 
In this paper, we introduce \textit{tail batching}, a novel rollout scheduling strategy for synchronous RL that systematically consolidates prompts leading to long-tail responses into a small subset of rollout steps (\textit{long rounds}), while ensuring that the majority of steps (\textit{short rounds}) involve only balanced, short rollouts. By excluding long responses from short rounds and rescheduling them into a few designated long rounds, tail batching effectively reduces GPU idle time during rollouts and significantly accelerates RL training without sacrificing accuracy. We present \SystemName{}, a system that fully harnesses the benefits of tail batching through holistic optimizations across all three RL stages: elastic parallelism adaptation for rollout, dynamic resource allocation and scheduling for reward, and stream-based training. Empirical results show that \SystemName{} 
achieves a $2.03\times$-$2.56\times$ end-to-end training time reduction compared to veRL~\cite{verl}, and up to $2.24\times$ speedup compared to RLHFuse~\cite{rlhfuse} for the Qwen2.5 family of LLMs on up to 128 H800 GPUs.

\end{abstract}

%% file: contents/1introduction.tex
%!TEX root=../main.tex
\section{Introduction}
\label{sec:introduction}

Advanced Large Language Models (LLMs)~\cite
{openaio4,qwq32,deepseekr1,seed-thinking} critically rely on Reinforcement
Learning (RL) post-training to enhance reasoning capabilities in complex
tasks, such as mathematics~\cite{maxwell_jia_2024_aime_dataset}, code generation~\cite
{openr1_codeforces_dataset}, and tool use~\cite
{pan2024trainingsoftwareengineeringagents,verl-agent}. The standard RL
post-training workflow for LLM reasoning models comprises repeated cycles
across three stages~\cite{shao2024deepseekmath,deepseekr1}: \textit
{rollout}, \textit{reward}, and \textit{training}. In the rollout stage, the
actor LLM generates responses for a batch of input prompts. These responses
are subsequently evaluated in the reward stage using various strategies, such as 
sandbox execution for coding tasks, rule-based logic
for mathematical problems, and LLM-as-a-Judge~\cite
{son2024llmasajudgerewardmodel} for nuanced tasks including human alignment. In
the final training stage, the actor LLM’s weights are updated based on the computed
reward signal, optionally with a reference LLM to ensure training stability.

To maximize model performance, LLM practitioners often employ \emph
{synchronous on-policy} RL post-training to guarantee that responses in the rollout
stage are always generated by the most recently updated actor LLM. This is
achieved by enforcing a synchronization barrier between the rollout stage and the
training stage~\cite{hybridflow,realhf,openrlhf,distflow}, as illustrated in Figure~\ref{subfig:sync_rl}. However, this
synchronization requirement frequently results in severe pipeline bubbles,
especially during rollout stages, which account for around 70\% of
the total training time in our experiments (see Table~\ref
{tab:step_time_breakdown}). Notably, rollout batches typically exhibit
a \emph{long-tail distribution} in response length, with the longest response
$25\times$-$32\times$ longer than the medium (see Figure~\ref
 {subfig:length-dist}). This imbalance leads to prolonged idle periods on
 GPUs generating short responses, as these devices need to idle wait until the
 entire batch are completed.

A common approach to mitigating idle bubbles is to overlap the long rollout
stage with reward computation (e.g., ROLL~\cite{roll} and MiMO~\cite
{MiMo}) and reference model inference (e.g., RLHFuse~\cite{rlhfuse}).
However, in LLM reasoning
post-training, the combined computation for reward evaluation and reference model inference typically accounts for less than 15\% of total training time (see Table~\ref
{tab:step_time_breakdown})---insufficient to fill idle
bubbles during long rollout periods.

Many recent RL systems have explored relaxing synchronization constraints for
more aggressive stage overlap.
One common solution is the ``one-off'' pipeline,
adopted by DeepScaler~\cite{deepscaler2025}, StreamRL~\cite{StreamRL}, and
AsyncFlow~\cite{asyncflow}, wherein rollouts generated in a previous step are
used for subsequent training. Some frameworks, such as AReaL~\cite
{areal}, even adopt fully asynchronous RL training that continuously performs
rollouts and training in parallel. Although these approaches effectively reduce
idle bubbles, they often compromise model
accuracy because long rollouts are produced with stale model weights relative
to short responses. As a result, many RL researchers and practitioners 
remain hesitant to adopt asynchronous training for LLM post-training.

In this paper, we propose \emph{tail batching}, a novel prompt scheduling
strategy designed for on-policy RL training that effectively mitigates GPU bubbles
induced by long-tail rollouts. Empirically, we observe that within a rollout
batch, only a small subset of prompts produce exceedingly long responses that
stall the entire batch. Our key idea is to reorder training samples by
consolidating these \emph{tail prompts} into a few \emph{designated} rollout
steps, referred to as \emph{long rounds}, while ensuring that the majority of
rollout steps (\emph{short rounds}) are composed of balanced, short responses,
thereby reducing idle bubbles in GPU utilization. Importantly, because tail batching
alters only the order of training samples, it preserves training accuracy, 
as indicated by recent algorithmic research~\cite{SPEEDRL, MoPPS, DAPO}.

To implement this approach, we present \SystemName{}, a system engineered to
unlock the full potential of tail batching for on-policy RL
training. \SystemName{} initiates rollout in a \emph{short round} to sample
$P_0$ prompts, each producing $R_0$ responses. To collect balanced, short
responses, \SystemName{} employs \emph{speculative execution} for both
prompts and responses in a short round: it launches more than $P_0$ prompts
but retains only the first $P_0$ that finish; each prompt produces more than
$R_0$ responses, finishing after the first $R_0$ complete. Prompts that
generate long responses and are excluded from a short round are deferred into
a long-prompt queue. Once the queue accumulates $P_0$ such
prompts, \SystemName{} batch-executes them in a dedicated \emph{long round},
without speculative execution.

\SystemName{} further introduces three system-level optimizations, each addressing
a bottleneck in a distinct stage of the RL training pipeline.
 First, we design a \emph{parallelism planner} that adaptively configures
 parallelization strategies during rollout. Compared to long rounds, short
 rounds impose higher GPU memory pressure because speculative execution launches more
 concurrent requests. As training proceeds, response length distributions 
 change significantly~\cite{deepseekr1}. A fixed parallelization
 scheme cannot accommodate this variability. To address this,
 \SystemName{} profiles memory footprint across different batch sizes
 and sequence lengths, then selects the best tensor parallelism (TP) configuration
 for each training step based on these profiles and the online response length distribution.
 This dynamic TP configuration quickly adapts to workload changes over RL training, 
 reducing rollout latency by up to 21.9\% in our evaluation (see \S\ref{sub:impact-parallel-planner}).

Second, as rollout cost reduces, reward computation can
become a bottleneck, particularly for code execution and
LLM-based judging tasks. To address this, \SystemName{} introduces a \emph
{reward scheduler} that performs asynchronous, per-sample reward computation.
It pipelines reward evaluation in parallel with ongoing rollouts to hide the
reward overhead, while dynamically adjusting the compute budget for each
reward task based on workload characteristics, such as adjusting sandbox
timeouts for code execution or dynamically sharing GPUs for judge models. This approach
substantially reduces reward and further speeds up the end-to-end latency for an average of 23.9\% in our evaluation (see \S\ref{sub:impact-reward-scheduler}).

Third, \SystemName{} implements a \emph{stream trainer} that overlaps rollout and training
to further reduce GPU idle time. As rollouts progress, especially in long
rounds, some GPUs may become idle if their assigned requests complete early. To
harvest these idle GPUs, the stream trainer opportunistically initiates training as soon
as a partial set of completed prompts is available, while scaling down the
GPUs dedicated to rollout. It uses a heuristic algorithm to decide
when and which GPUs are reassigned, ensuring minimal disruption to rollout. 
Completed prompts are asynchronously streamed to the training
stage, allowing gradient computation and optional reference logit evaluation
to proceed in parallel with ongoing rollouts. To ensure the gradients computed
are consistent with those in synchronous on-policy training, \SystemName
{} adjusts the loss scales and defers gradient updates until the
streaming concludes. This design minimizes idle bubbles across stages while
maintaining on-policy training semantics.

We implemented \SystemName{} in 6.6k lines of Python code atop ROLL~\cite{roll}\footnote{The code will be released upon publication. }. We train models from the Qwen2.5 family (7B–32B)~\cite{qwen25} with real-world datasets~\cite{he2025deepmath,xu2025kodcode} on a cluster of 128 H800 GPUs using \SystemName{}. Evaluation results show that \SystemName{} substantially outperforms state-of-the-art RL systems, achieving $2.03\times$-$2.56\times$ end-to-end training speedup over veRL~\cite{verl} and up to $2.24\times$ speedup compared to RLHFuse~\cite{rlhfuse}.

%% file: contents/2background.tex
\begin{figure}[tb]
    \centering
    \begin{subfigure}{0.49\textwidth}
        \centering
        \includegraphics[width=\linewidth]{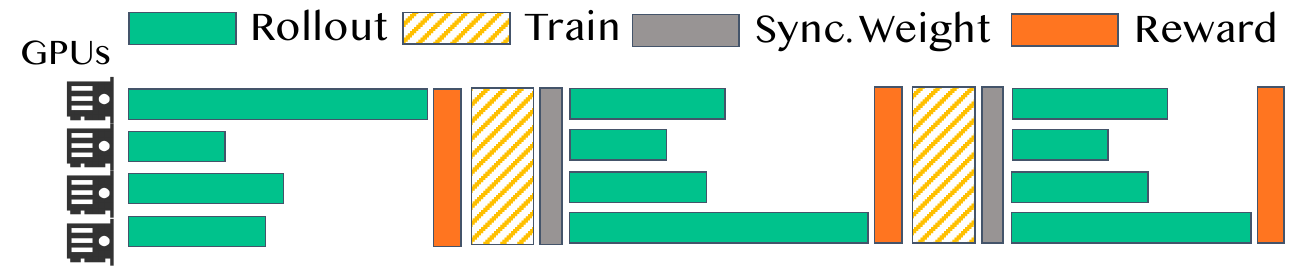}
        \caption{Synchronous RL Post-Training.}
        \label{subfig:sync_rl}
    \end{subfigure}
    \begin{subfigure}{0.49\textwidth}
        \centering
        \includegraphics[width=\linewidth]{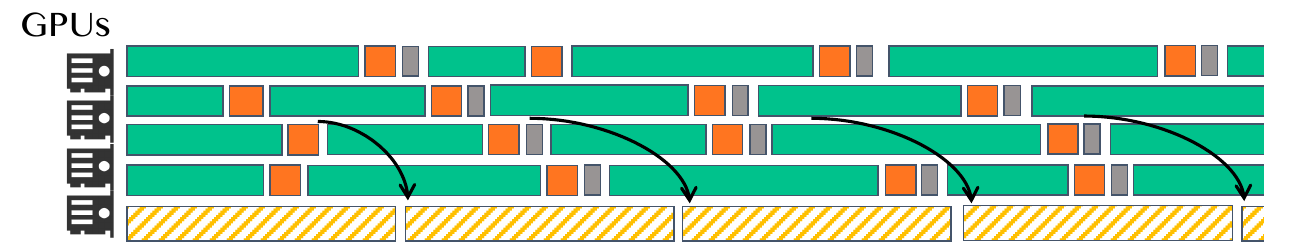}
        \caption{Asynchronous RL Post-Training.}
        \label{subfig:async_rl}
    \end{subfigure}
    \caption{
        Execution workflows of synchronous and asynchronous RL post-training.
        % 
        % (a) In synchronous-RL training, the actor LLM training utilizes the samples generated by the latest actor LLM.
        % (b) In asynchronous-RL training, the actor LLM can be trained using stale data generated in previous steps.
        % One popular asynchronous-RL training paradigm is the one-off pipeline~\cite{deepscaler2025}, which allows the train stage to utilize the data from the last step.
    }
    \label{fig:rl_workflow}
\end{figure}

\section{Background and Motivation}
\label{sec:background}

\subsection{RL for LLM Post-Training}
\label{subsec:bg-rl-for-llm}

% \noindent\textbf{Models and Workflow.}

Reinforcement Learning (RL) has become a pivotal technique for post-training
LLMs. Recent advances show that RL algorithms such as GRPO~\cite
{shao2024deepseekmath} are highly effective in enhancing reasoning capabilities
across various domains. An RL post-training workflow typically orchestrates
multiple models with distinct roles. The \textit{actor LLM} generates
responses to input prompts and serves as the primary model being optimized.
The \textit{reward LLM} evaluates each response and outputs a scalar reward
signal, which can be derived from heterogeneous sources such as sandbox
execution for code, rule-based logic for mathematics, or through 
LLM-as-a-Judge~\cite{son2024llmasajudgerewardmodel} for alignment tasks. To
further stabilize optimization, a \emph{reference LLM} is often introduced as
a regularizer. Overall, the workflow comprises three stages: (1) \emph
{rollout}, where the actor LLM is given $P_0$
prompts and produces $R_0$ responses for each prompt; 
(2) \emph{reward}, where the generated responses are evaluated by
corresponding reward workers; and (3) \emph{training}, where the actor LLM
updates its parameters based on the computed rewards, optionally constrained
by the reference model to mitigate gradient instability.

To maximize model performance, RL post-training is usually performed in a
\emph{synchronous} setting, known as \emph{on-policy training}. In this
 setting, rollout must complete before training begins, and the actor's
 weights are updated only after training concludes (see Figure~\ref
 {subfig:sync_rl}). This synchronization requirement ensures that all
 responses are generated using the most recent model parameters, thereby
 stabilizing training and improving reliability across tasks~\cite
 {sync_rl_stable}. However, it often results in severe pipeline
 bubbles and low utilization, especially in the rollout stage, as shown in our characterization study.

\begin{figure}[tb]
    \centering
    \begin{subfigure}{0.23\textwidth}
        \centering
        \includegraphics[width=\linewidth]{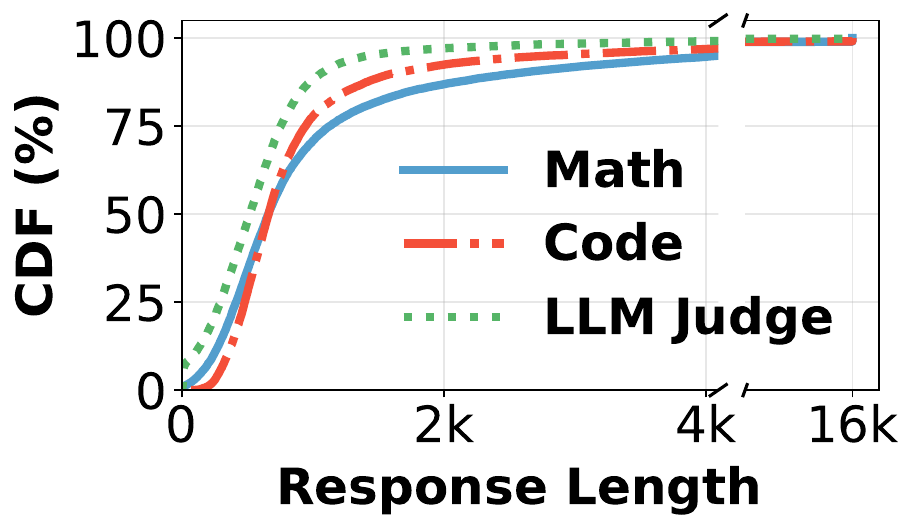}
        \caption{Length Distribution.}
        \label{subfig:length-dist}
    \end{subfigure}
        \begin{subfigure}{0.23\textwidth}
        \centering
        \includegraphics[width=\linewidth]{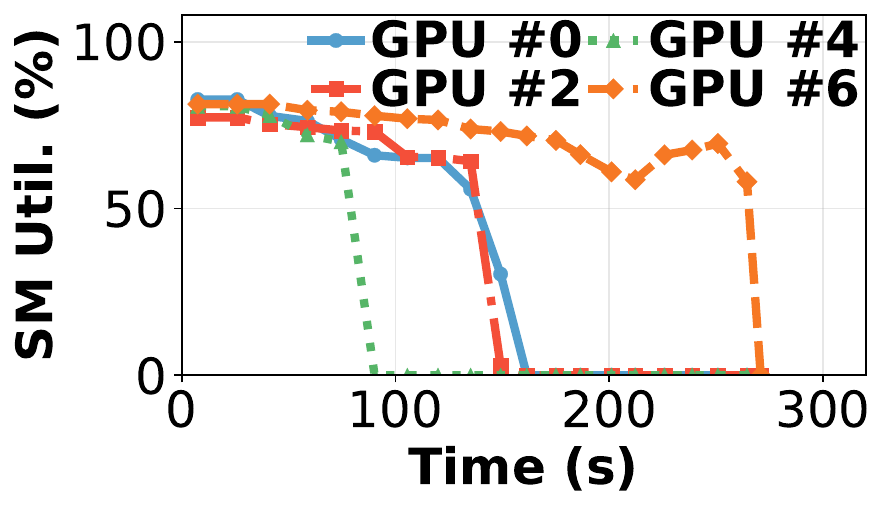}
        \caption{SM Utilization.}
        \label{fig:gpu-util}
    \end{subfigure}
    \caption{
        Characterizing rollout stage:
        (a) responses across three tasks exhibit a long-tail distribution;
        (b) long-tail rollouts create prolonged GPU bubbles in the rollout stage.
    }
    \label{fig:motivation-lpt-characterization-inference}
\end{figure}

% \begin{figure}[tb]
% \centering
% \includegraphics[width=.5\linewidth]{figs/length_cdf_full.pdf}
% \caption{Length Distribution of multiple training datasets.}
% \label{fig:length_distribution}
% \end{figure}

\begin{table}[tb]
    \centering
    % \scriptsize
    \caption{Time breakdown of RL post-training. We train 14B models with a maximum length of 16k using veRL~\cite{verl} and GRPO~\cite{shao2024deepseekmath} with real-world datasets~\cite{he2025deepmath,xu2025kodcode} in three tasks.}
    \resizebox{0.95\linewidth}{!}{
    % \begin{tabularx}{0.\linewidth}{lccc}
    \begin{tabularx}{\linewidth}{l *{3}{>{\centering\arraybackslash}X}}
    \toprule
        \textbf{Task} & \textbf{Rollout~~} & \textbf{Reward} & \textbf{Training}\\ 
        \midrule
        Math & 72\% & 5\% & 23\% \\ 
        % \midrule
        Code & 66\% & 13\% & 21\% \\ 
        % \midrule
        LLM-as-a-Judge & 71\% & 7\% & 22\% \\ 
        \bottomrule
    \end{tabularx}}
    \label{tab:step_time_breakdown}
\end{table}

\subsection{Characterization of RL Post-Training}
\label{subsec:character-rl-training}

We characterize RL post-training workloads using the Qwen2.5‑14B model~\cite
{qwen25}, configured with a maximum response length of 16k tokens, a batch
size of 128, and a group size of 8 under the GRPO algorithm~\cite
{shao2024deepseekmath}. We run math, code, and LLM-as-a-Judge tasks with
real-world datasets~\cite{he2025deepmath,xu2025kodcode} using veRL~\cite
{verl} on 32 H800 GPUs. For the LLM‑as‑a‑Judge experiments, we employ a
7B‑parameter judge model. 

\noindent\textbf{The Rollout Bottleneck.}
Table~\ref{tab:step_time_breakdown} reports the stage-wise latency
distribution for RL training under these settings. The rollout stage
dominates runtime, accounting for approximately 70\% of each training step across
three tasks. The reward stage
contributes a smaller fraction (5\%–13\%), while the training stage accounts
for 21\%–23\% of the total step time. 
% Reference model inference, typically grouped with training, adds less than 5\% to the end-to-end latency. 
These results underscore that rollout is the primary bottleneck in RL post-training. 

% Notably, reward evaluation remains non-trivia, particularly for coding tasks, 
% where sandbox execution can introduce noticeable delays. We next examine rollout 
% and reward stages in greater detail.

% Table~\ref{tab:step_time_breakdown} presents a performance breakdown of RL training with a maximum response length of 32K tokens. The rollout stage dominates the execution time, accounting for approximately 59\%, 60\%, and 62\% of each RL training step for the math, code, and LLM-judge tasks, respectively. Reward computation contributes roughly 3\%–14\% across tasks, while the training stage accounts for 26\%-38\% of the total end-to-end training time. Overall, the rollout stage dominates the RL training overhead. Contrary to the stereotype, the cost of reward evaluation is non-negligible, especially for the code task. We provide further analysis for rollout and reward stage as follows. 

% son2024llmasajudgerewardmodel JUDGE

\noindent\textbf{Long-Tail Response and GPU Bubbles.}
A key source of inefficiency in RL post-training lies in the \emph
{highly skewed distribution} of response lengths. As shown in Figure~\ref
{subfig:length-dist}, responses across math, code, and LLM-as-a-Judge tasks
exhibit a pronounced \emph{long tail} distribution: while most responses are short to
moderate in length, with the 75th percentile (P75) ranging between 755 and
1.1k tokens, the longest responses can extend up to 16k tokens.

The presence of long-tail responses results in poor GPU utilization under
synchronous rollout. Because all GPUs must wait for the longest responses to
complete, devices assigned shorter requests become idle, creating prolonged
``bubbles'' of wasted cycles. Figure~\ref{fig:gpu-util} illustrates this
problem by reporting SM utilization of even-indexed GPUs on a server
with tensor parallelism configured to two. Utilization peaks near 80\% at the start of
rollout but never reaches full saturation, as LLM decoding is inherently
memory-bound. Once short responses finish, utilization of corresponding GPUs
quickly drops to zero, with idle periods lasting until the entire batch
completes. Given that rollout is already the dominant
contributor to training latency, such inefficiency significantly stalls the
entire pipeline, a problem widely reported in the 
literature~\cite{RhymeRL,rlhfuse,asyncflow,areal}.

% as those
% assigned for short responses must wait until the longest to complete under
% the synchronization requirement, whose length can be 25$\times$ to 32$\times$
% of the medium (see Figure~\ref{subfig:length-dist}). To illustrate this
% problem, we depict in Figure~\ref{fig:gpu-util} the SM utilization of
% different GPUs at various time points during a rollout stage in a GPU server.
% We report results only for GPUs with even indices because the TP size is
% configured to two. The GPU utilization peaks at the beginning of rollout
% (approximately 80\%), but still cannot reach full saturation because LLM
% decoding is inherently memory-bound. As assigned responses complete, the GPU
% utilization drops sharply to zero, and the devices remain idle until the
% entire batch complete. This leads to prolonged bubbles, up to 67\% of the
% rollout stage on GPU \#4 as shown in Figure~\ref{fig:gpu-util}.
% As rollout dominates the duration of RL post-training, its inefficiency 
% substantially delays the entire pipeline, a problem widely reported in
% existing works~\cite{RhymeRL,rlhfuse,asyncflow}.

% Overall, the long-tail rollouts results in GPU underutilization and causes substantial waste of computational resources, emphasizing the importance of tailored system optimizations.

\subsection{Existing Solutions and Limitations}
\label{subsec:existing_solutions_and_limitations}

% Existing systems employ two approaches to reducing GPU bubbles caused by the long-tail rollouts in
% RL post-training.

Prior efforts to mitigate GPU bubbles in RL post-training generally fall into
two categories: \emph{stage overlap under synchronization constraints} and \emph{relaxed
synchronization}.

% To settle the low resource utilization and the increased rollout time caused by the long-tail response in RL post-training, existing systems either adopt an overlap-based approach to pipeline other stages with the long-tail response generation, or introduce an asynchronous RL post-training paradigm to relax the tight synchronization requirements of the train stage and the rollout stage to enable a decoupled pipeline.

\noindent\textbf{Stage Overlap under Synchronization Constraints.}
This approach seeks to improve resource utilization by pipelining the
long-tail rollout with the execution of other stages before the
synchronization barrier. For example, RLHFuse~\cite{rlhfuse} overlaps
rollout with reward computation and reference model inference, while
frameworks like ROLL~\cite{roll} and MiMO~\cite{MiMo} overlap the reward
computation of each completed response with the ongoing rollout stage to
enable \textit
{asynchronous reward computation}. While these designs reduce idle bubbles
to some extent, they do not fundamentally address the long-tail responses
that dominate rollout time with only modest performance improvements 
(see \S\ref{subsec:e2e-eval}). Moreover, as response lengths in RL
post-training continue to grow~\cite{Polaris2025}, the relative
contribution of reward and reference inference diminishes (typically less
than 15\% of step runtime as reported in Table~\ref
{tab:step_time_breakdown}), leaving insufficient work to mask the prolonged
idle bubbles, even under ideal overlap.

% RLHFuse~\cite{rlhfuse} pipelines the reward and reference model inference with the long-tail rollout.
% ROLL~\cite{roll} and MiMo~\cite{MiMo} adopt \textit{asynchronous reward computation} to overlap the reward computation of each completed response with the ongoing rollout stage.
% These works tackles resource idleness by overlapping; however, these approaches fail to fundamentally reduce the long-tail responses and only achieve a marginal step time reduction.
% With the ever-growing response length in RL post-training for reasoning LLMs (e.g., for more than 32K~\cite{Polaris2025}), overlapping-based approaches fall short in settling the long-tail rollout.

% these methods still suffer from the long-tail response generation.

\noindent\textbf{Relaxed Synchronization.}
A second line of work adopts a more aggressive strategy by relaxing the strict
synchronization barriers between rollout and training, as illustrated in
Figure~\ref{subfig:async_rl}. For example, Kimi~\cite
{kimiteam2025kimik15} introduces \textit{partial rollout} by truncating the
long-tail responses and preserving generated tokens to continue rollouts in
subsequent steps. StreamRL~\cite{StreamRL}, AsyncFlow~\cite{asyncflow}, and
RhymeRL\footnote{RhymeRL's HistoPipe scheduling requires one-step
off-policyness (see Figure~10 and evaluation in \S7.3 of~\cite
{RhymeRL}).}~\cite{RhymeRL} allow a one-step staleness, enabling the training
stage to proceed with slightly outdated rollouts in a \emph
{one-off pipeline}. Pushing further, AReaL~\cite{areal} introduces \emph
{fully asynchronous} RL training, in which rollout and training are
completely decoupled, and updates may rely on samples generated many steps
earlier. These methods are effective in reducing GPU bubbles, but they
introduce a fundamental trade-off: by relaxing synchronization, they
compromise the on-policy nature of training, often leading to degraded
accuracy and reduced stability.

In summary, existing solutions either provide only marginal improvements by
overlapping non-bottleneck stages with rollout, or sacrifice on-policy
guarantees by relaxing synchronization. Neither approaches can eliminate the
inefficiency introduced by long-tail rollouts while preserving the accuracy
and stability of synchronous RL training.

%% file: contents/3designprinciples.tex
\section{Tail Batching}
\label{sec:tail_batching}

\begin{figure}[tb]
\centering
\includegraphics[width=\linewidth]{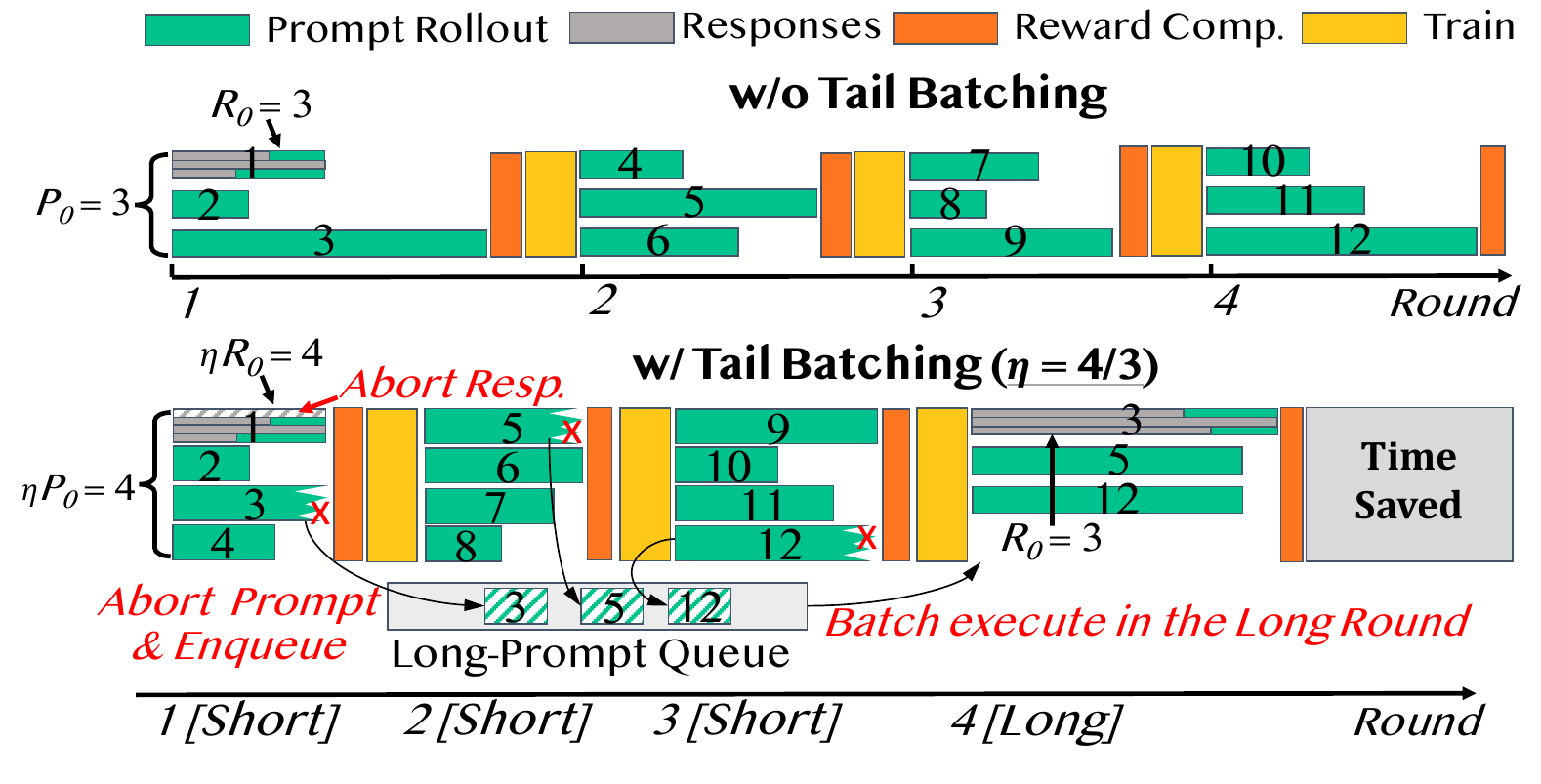}
\vspace{-15pt}
\caption{Illustration of Tail Batching.}
\label{fig:prompt_squeezer}
\end{figure}

In this section, we introduce \emph{tail batching}, a novel prompt scheduling
strategy that fundamentally alleviates imbalanced response lengths to reduce
GPU bubbles while preserving on-policy RL training semantics without accuracy
loss.

\noindent\textbf{Tail Batching.} 
GPU bubbles arise primarily from a small subset of prompts that generate
disproportionately long responses. A naive approach would be to exclude such
long-trail responses from rollout batches. However, this approach introduces
two critical issues: (\textbf{P1}) rollout stages may fall short of the required
number of prompts or responses that constitute an effective batch; and
(\textbf{P2}) systematically excluding long prompts distorts the training sample
distribution, potentially harming model performance. Tail batching addresses
these problems with two key techniques. 

To address \textbf{P1}, tail batching leverages \emph{speculative execution}~\cite
{spark} by over-provisioning requests while selectively retaining only the
fastest completions. Under the GRPO algorithm~\cite{shao2024deepseekmath},
each rollout step requires sampling $P_0$ prompts, each with
$R_0$ responses (Figure~\ref{fig:prompt_squeezer}-top).
Instead of launching exactly $P_0$ prompts,
tail batching starts more and admits only the first
$P_0$ to complete. Similarly, each prompt produces more than $R_0$ responses,
but only the first $R_0$ are retained. This ``race-to-completion''
speculation naturally filters out long responses, yielding 
balanced, shorter batches that minimize idle bubbles while
preserving the required batch size.

To address \textbf{P2}, tail batching guarantees that no prompt is permanently
excluded. As shown in Figure~\ref{fig:prompt_squeezer}-bottom, 
prompts aborted during speculative execution are added to a \emph
{long-prompt queue}. Once the queue reaches size $P_0$, these prompts are
batch-scheduled in a dedicated \emph{long round}, where speculative execution
is \emph{disabled} to allow full-length responses to be generated. Because
such prompts are rare, long rounds occur infrequently and are interleaved
with frequent \emph{short rounds} composed of balanced responses. This design
ensures that all prompts are eventually included, while the majority of
rollout steps remain efficient.

\noindent\textbf{Training Accuracy.}
From a statistical perspective, tail batching only reorders short and long
prompts into separate rounds without altering the underlying sample
distribution or relaxing synchronization. Prior studies show that changes in
training order do not degrade model accuracy~\cite
{DBLP:conf/eacl/ChangYD21, Bengio2009, DBLP:conf/corl/FlorensaHWZA17}. In
fact, several recent RL post-training algorithms explicitly explore prompt
reordering as a means of efficiency~\cite{SPEEDRL, MoPPS, DAPO}. Our
empirical results further validate this claim: as shown in Figure~\ref
{fig:e2e_score}, tail batching achieves accuracy curves nearly identical to those of
standard synchronous RL training.

\begin{figure}[tb]
    \centering
    \begin{subfigure}{0.23\textwidth}
        \centering
        \includegraphics[width=\linewidth]{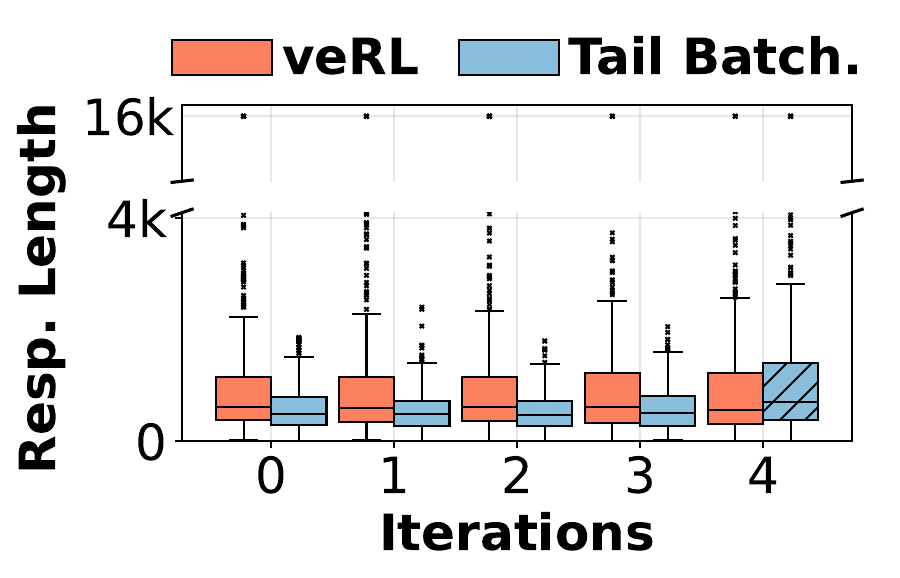}
        \caption{Length Distribution.}
        \label{subfig:box_length_distribution}
    \end{subfigure}
    \begin{subfigure}{0.23\textwidth}
        \centering
        \includegraphics[width=\linewidth]{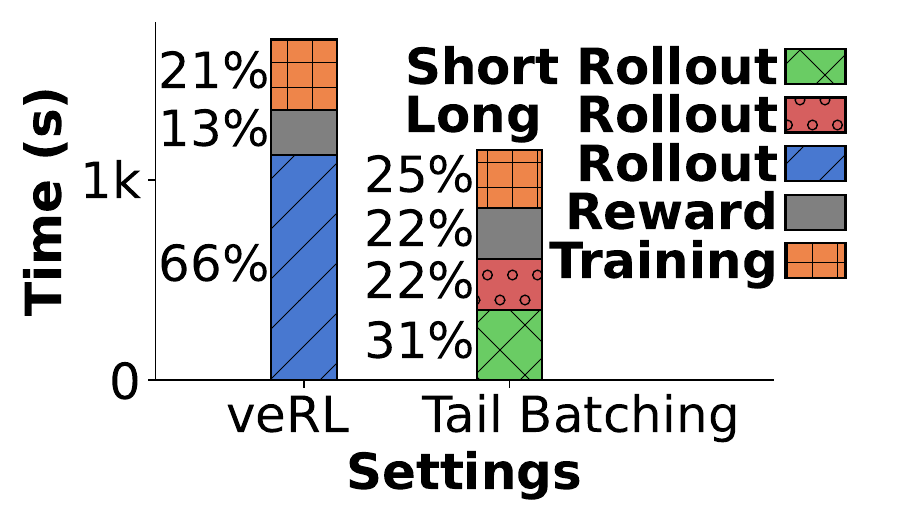}
        \caption{Training Time Breakdown.}
        \label{fig:bkd_bar_sqeezer}
    \end{subfigure}
    \caption{
        Tail batching vs. the baseline veRL when training a Qwen2.5-14B model on the code dataset~\cite{xu2025kodcode}.
        (a) Box plot of response length distribution 
        across five rounds, where the hatched box is a long round under tail batching. 
        Whiskers measure 1.5 IQR.
        (b) Training time breakdown, where
        the total time is a cumulation of 5 consecutive steps, a 
        full period comprising four short rounds and one long round.
    }
    \label{fig:squeezer_motiv}
\end{figure}

\noindent\textbf{Rollout Efficiency.}
We empirically validate tail batching's effectiveness in enhancing rollout
efficiency by training a Qwen2.5‑14B model~\cite{qwen25} on the code
dataset~\cite{xu2025kodcode} under the same settings described in \S\ref
{subsec:character-rl-training}. The speculation factor is set to $\eta = 1.25$, meaning that in each short round, the actor LLM speculatively launches
$\eta P_0$ prompts, each generating $\eta R_0$ responses, while
accepting only the first $P_0 \times R_0$ completions.
Figure~\ref{subfig:box_length_distribution} compares the response length
distribution over five training steps, with and without tail batching.
Compared to the baseline approach (veRL~\cite{verl}), tail batching yields shorter 
and more balanced responses in the first four steps
(short rounds), reducing the maximum response length by up to $8.9\times$. Prompts producing long responses are deferred to the fifth step
(long round), where outputs are generally longer than the baseline but capped at the same
maximum of 16k tokens.
This reorganization substantially reduces rollout costs and 
shortens end-to-end training time by $1.48\times$ (Table~\ref{tab:ablation_e2e}).

Figure~\ref{fig:bkd_bar_sqeezer} further breaks down the training time across
stages. With rollout overhead mitigated by tail batching, the
relative contributions of the reward and training stages become more
pronounced. Moreover, short and long rounds exhibit drastically different
resource usage profiles, implying that a uniform rollout strategy is
suboptimal. These findings motivate the system-level
optimizations tailored to each stage of the RL pipeline, which we develop next.

%% file: contents/4systemdesign.tex
%!TEX root=../main.tex
\section{\SystemName{} System Design} 
\label{sec:system_design}

\begin{figure}[tb]
\centering
\includegraphics[width=\linewidth]{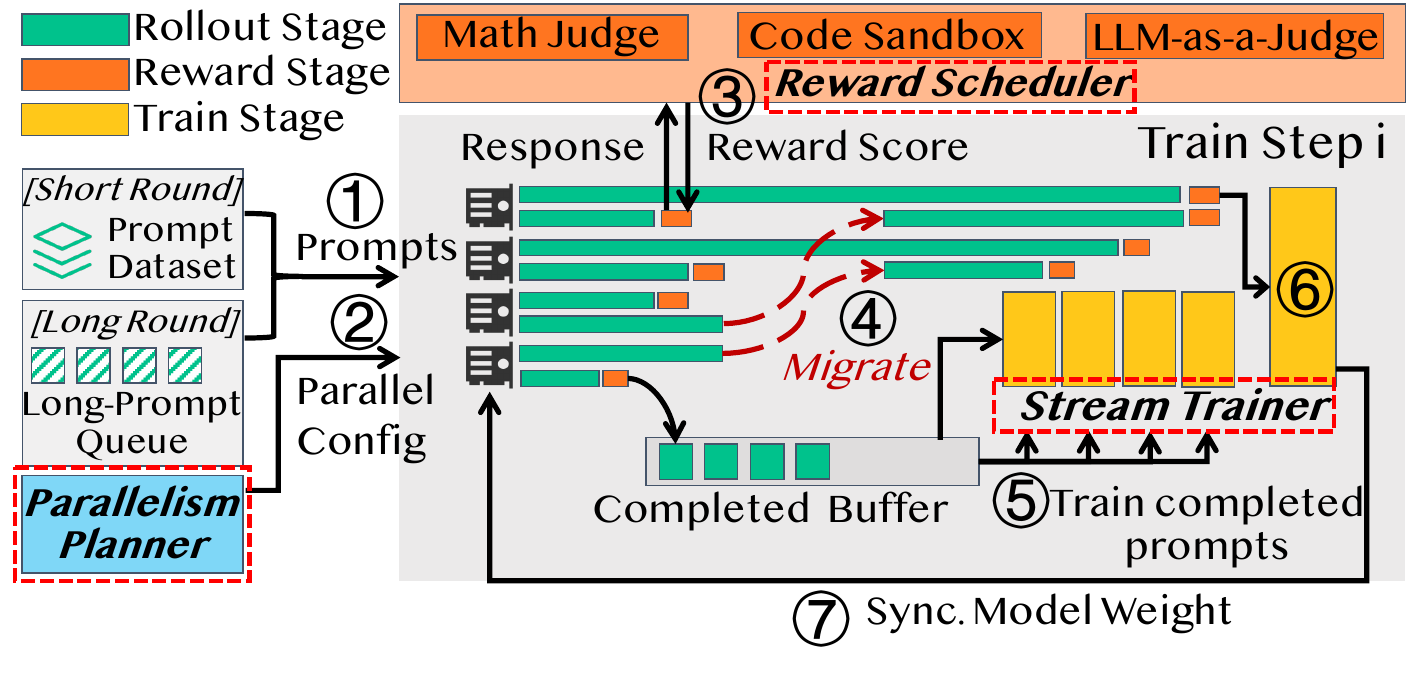}
\vspace{-18pt}
\caption{System overview and workflow of \SystemName{}.}
\label{fig:system_overview}
\end{figure}

In this section, we present \SystemName{}, an efficient on-policy RL system
engineered to fully realize the benefits of tail batching through a holistic design.
We begin with a system overview, then provide detailed descriptions of each component.

\subsection{System Overview}
\label{subsec:system_overview}

% \noindent \textbf{Key Components.}
\SystemName{} incorporates three key components: the parallelism planner, reward
 scheduler, and stream trainer, each addressing a distinct bottleneck in 
 rollout, reward, and training stages.

\noindent \textbf{Parallelism Planner.}
Short rounds, which employ speculative execution, create higher
GPU memory pressure than long rounds. A fixed tensor parallelism (TP)
configuration cannot adapt to the changing resource profiles across short and long
rounds, resulting in frequent KV cache preemption overhead and degraded efficiency.
\SystemName{} introduces a parallelism planner that dynamically profiles workloads and selects optimal TP configurations each step to cut rollout overhead.

\noindent \textbf{Reward Scheduler.}
With rollout costs reduced, reward computation becomes more pronounced
(Figure~\ref{fig:bkd_bar_sqeezer}). To prevent it from
becoming the new bottleneck, \SystemName{} employs a reward scheduler that
pipelines reward computation in parallel with rollouts while 
dynamically budgeting compute for each sample evaluation, effectively reducing
its overhead.

\noindent \textbf{Stream Trainer.}
In long rounds where speculative execution is disabled, imbalanced responses lead to pronounced GPU bubbles (Figure~\ref{subfig:box_length_distribution}). The stream trainer advances prior stage-overlapping approaches~\cite{rlhfuse,roll} by introducing a more fine-grained overlap between rollout and training: completed prompts are streamed into training immediately, while idle GPUs are reassigned from rollout to gradient computation. To maintain on-policy semantics, the stream trainer carefully scales gradients and defers weight updates until the full rollout completes, preserving accuracy while reducing idle time.

\noindent\textbf{Workflow.} 
\SystemName{} operates in two phases: an \emph{offline profiling phase} and \emph{online
 execution phase}. In the \textbf{offline phase}, \SystemName{} benchmarks the actor
 LLM's prefilling and decoding throughput under different TP sizes, batch
 sizes, and sequence lengths. It also profiles the GPU memory footprint and
 runtime cost of the judge LLM across varying sequence lengths. These profiled
 results are used for guiding online decisions.

In the \textbf{online phase}, \SystemName{} orchestrates rollout, reward, and training
in a synchronous RL job (Figure~\ref{fig:system_overview}). 
\wcircle{1} During rollout, tail batching decides whether to apply speculative
 execution based on the size of the long-prompt queue. 
\wcircle{2} The parallelism planner then selects an optimal TP configuration
 by combining historical job loads with profiled performance data.
\wcircle{3} In parallel,
the reward scheduler overlaps evaluation with rollout and dynamically adjusts budgets for each reward task. 
\wcircle{4} Concurrently, the stream trainer monitors rollout progress to determine
 when to reassign GPUs from rollout to training.
\wcircle{5} As prompts complete, they are streamed into training for immediate gradient computation.
\wcircle{6} Once the full rollout completes, the stream trainer stops
 streaming, accumulates all computed gradients, and triggers a
 synchronized
 gradient computation and update across all available GPUs.
\wcircle{7} Finally, the updated actor weights are synchronized with the
 rollout stage before the next RL step begins.

\subsection{Parallelism Planner}
\label{subsec:parallel-planner}

\begin{figure}[tb]
    
    % 进一步微调与下方两图的垂直间距（可根据需要调整或去掉）
    % \vspace{-0.1em}
    
    % 底部：左右两张子图
    \begin{subfigure}{0.23\textwidth}
        \centering
        \includegraphics[width=\linewidth]{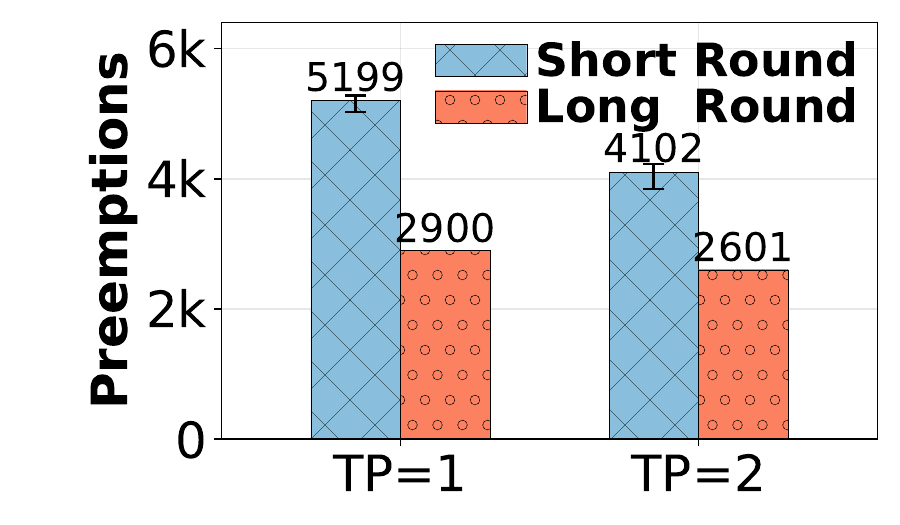}
        \caption{Preemption Count.}
        \label{fig:motiv_preempt_num}
    \end{subfigure}
    \hfill
    \begin{subfigure}{0.23\textwidth}
        \centering
        \includegraphics[width=\linewidth]{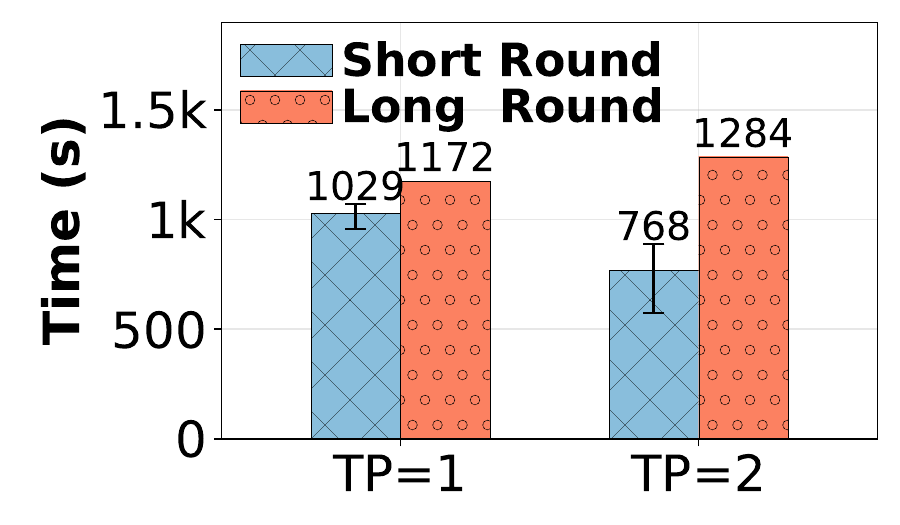}
        \caption{Rollout Time.}
        \label{fig:motiv_preempt_rollout_time}
    \end{subfigure}
    
    \caption{
        Rollout performance when training Qwen3-8B/32k on eight H800 GPUs.
        (a) Preemption count in each step.
        (b) Rollout time of each step.
        The metric is collected in one consecutive period of 4 short rounds and one long round.
    }
    \label{fig:motiv_parallelism_planner_preemptions}
\end{figure}

\noindent\textbf{Short Rounds Create High Memory Pressure.} 
As described in \S\ref{sec:tail_batching}, tail batching increases the number of concurrent responses in short rounds, placing greater pressure on GPU memory. Existing LLM serving engines~\cite{vllm,sglang} typically alleviate memory pressure by preempting ongoing requests, i.e.,
swapping out their KV cache to free GPU memory for others. A high preemption count thus indicates extensive memory contention. Figure~\ref{fig:motiv_preempt_num} shows that when training Qwen3-8B with a 32k response length and a batch size of 128, short rounds incur up to 1.79$\times$ more preemptions than long rounds, introducing substantial computational overhead. %  and delaying the decoding progress.

\noindent\textbf{Increasing TP Alleviates GPU Memory Pressure.}  
Tensor Parallelism (TP) is a standard technique to alleviate GPU memory
pressure. As shown in Figure~\ref{fig:motiv_preempt_num}, configuring a
larger TP size partitions model weights across more GPUs, freeing memory for
KV cache and cutting preemption counts by 21.1\% in short rounds and 10.3\%
in long rounds. This additional KV cache capacity alleviates memory
contention and shortens rollout latency.  As shown in Figure~\ref
{fig:motiv_preempt_rollout_time}, with TP=1, the rollout time in a short
round is 87\% of that in a long round, negating the gains of tail batching;
increasing TP size to 2 reduces short-round duration by 25\%.  
However, a larger TP size also introduces communication overhead, which dominates 
in long rounds where rollout is bound by long-tail responses, eroding performance.

% LLM serving~\cite{vllm,sglang} and

\noindent\textbf{Adaptive TP Selection.}
Most RL training frameworks~\cite{verl,openrlhf} adopt a fixed
parallelism configuration, which our analysis shows is
inefficient for tail batching (Figure~\ref
{fig:motiv_parallelism_planner_preemptions}). Optimal TP sizes differ between
short and long rounds, necessitating dynamic adaptation. \SystemName
{} introduces a \emph{parallelism planner} to reconfigure TP sizes on
a per-step basis with negligible overhead. In the offline phase, the planner
profiles optimal TP sizes without tail batching and uses them as default
configurations at the beginning of the training. It then keeps track of the
preemption counts and adapts the TP sizes accordingly using a lightweight
heuristic: a sudden rise in preemptions (e.g., $>1.05\times$) triggers an
increase in TP (doubling the size), while sustained zero preemptions 
across four steps trigger a decrease (halving the size). 
To limit cross-node communication, TP groups are constrained within
a single GPU server.

\begin{figure}[tb]
    \centering
    \begin{subfigure}{0.23\textwidth}
        \centering
        \includegraphics[width=\linewidth]{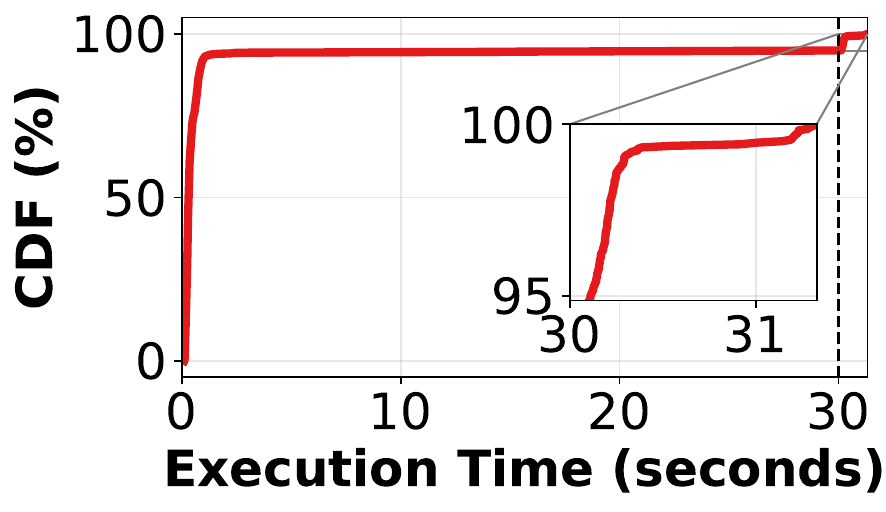}
        \caption{Code Sandbox.}
        \label{fig:code-sandbox}
    \end{subfigure}
        \begin{subfigure}{0.23\textwidth}
        \centering
        \includegraphics[width=\linewidth]{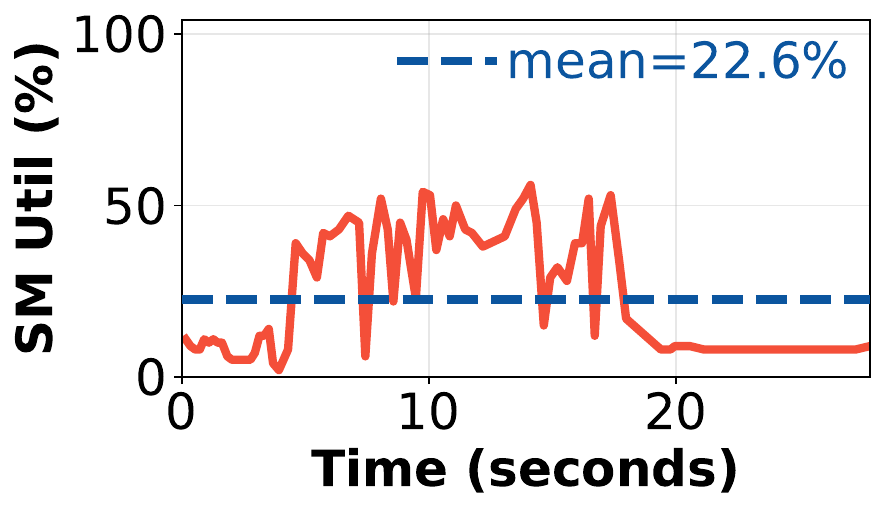}
        \caption{LLM-as-a-Judge.}
        \label{fig:llm-as-a-judge}
    \end{subfigure}
    \caption{
Reward computation introduces a non-negligible overhead with tail batching. (a) The distribution of sandbox execution time for code responses. (b) The SM utilization on the GPU allocated to a 7B-parameter judge LLM over time.}
    \label{fig:motivation-lpt-characterization-reward}
\end{figure}

\subsection{Reward Scheduler}
\label{subsec:reward-scheduler}

With rollout costs reduced, reward computation becomes a non-trivial
contributor to end-to-end latency (Figure~\ref{fig:bkd_bar_sqeezer}). To
mitigate this, \SystemName{} pipelines reward evaluation with rollout and
adaptively budgets compute for each task.

\noindent\textbf{Asynchronous Reward Computation.} 
In this method, reward evaluation is performed asynchronously: completed
responses are dispatched to reward workers in parallel with the ongoing rollout
stage, similar to ROLL~\cite{roll} and MiMo~\cite{MiMo}. This design overlaps
reward evaluation with rollout, partially hiding the overhead. However,
only relying on this design is insufficient to address
the potential bottleneck, especially for code sandbox execution and judge LLM
evaluation.

\noindent\textbf{Code Sandbox Execution.} 
In coding task, practitioners often impose a maximum execution timeout per
test case. For example, in our training experiments described in Figure~\ref
{fig:squeezer_motiv}, a 30-second timeout is enforced. Yet, as shown in
Figure~\ref{fig:code-sandbox}, around 5\% of prompts hit this timeout. These
prolonged executions, which ultimately yield zero reward, delays the entire
reward stage, stretching it to nearly 22\% of the total training time
(Figure~\ref{fig:bkd_bar_sqeezer}). Since many correct responses complete
much faster, a fixed timeout wastes substantial computation on doomed
samples.

\SystemName{} introduces an \emph{adaptive timeout
mechanism}. For each test case, it tracks the maximum execution time among
correct responses during training, denoted as $T_{\text{anchor}}$. When a new
response exceeds this threshold, sandbox execution is terminated early and a
zero reward is assigned. In view of the potential CPU contention during code
execution and to avoid overly aggressive cutoffs, the timeout is relaxed to

\begin{equation*}
  \displaystyle
  T_{\text{timeout}} = \min\big( \max(T_{\text{min}}, \lambda T_{\text{anchor}}), T_{\text{max}} \big),
\end{equation*}
where we empirically set $\lambda=1.5$, $T_{\text{min}}=2$s, and $T_{\text{max}}=30$s
to attain good performance. This design fast fails doom cases while 
preserving the evaluation of promising responses.

\noindent\textbf{LLM-as-a-Judge.} 
In asynchronous reward computation, RL systems often reserve a fixed number of
GPUs (e.g., 25\% of total GPUs) \emph{exclusively for the judge LLM} to avoid
interference with other workers. However, this strategy results in poor
utilization. As shown in Figure~\ref{fig:llm-as-a-judge}, when a 7B-parameter
judge LLM scores responses, its reserved GPU achieves only $\sim22.6\%$
average SM utilization. The inefficiency stems from the fact that the judge
typically processes small batches of responses, leaving much of the reserved
capacity idle.

To improve efficiency, \SystemName{} \emph{colocates the judge LLM with the
actor LLM} on the same GPU devices for concurrent execution. This design,
however, introduces two issues. First, rollout and reward evaluation now
share GPU resources, potentially interfering with one another. Nevertheless,
we observe that neither the actor LLM nor the judge LLM alone saturates GPU
SM utilization. To enable efficient sharing, \SystemName{} enables \emph
{Multi-Process Service} (MPS)~\cite{nvidia_mps}, which partitions
GPU resources at the warp level and allows both models to run concurrently
with minimal interference.

Second, hosting both models on the same GPU risks exhausting memory, as the
actor LLM already requires substantial space for its KV cache. To address
this, \SystemName{} introduces a \emph{layer‑wise pipeline scheme} that
reduces the memory footprint of the judge LLM. Inspired by prior work~\cite
{pipeswitch}, it offloads most layers of the judge LLM to host memory and streams
its parameters over PCIe in sync with activation computation on the GPU. Since
rollout rarely saturates PCIe bandwidth, this pipelined offloading imposes
little overhead. \SystemName{} dynamically adjusts the number of offloaded
layers to accommodate varying input sequence lengths, ensuring the judge LLM
fits within memory while maximizing utilization.

\begin{algorithm}[tb]
\caption{Stream Trainer}
\label{alg:stream-trainer}
\begin{algorithmic}[1]
\footnotesize
\State \textbf{Input:} Requests $R$, GPUs $G$
\Procedure{StreamTrainer}{$R, G$}
    \State $G_{\text{rollout}} \gets G$; \quad $G_{\text{train}} \gets \emptyset$
    \State $R_{\text{run}} \gets R$; \quad $R_{\text{comp}} \gets []$
    \State $scaled\_down \gets \text{false}$
    \State $\Delta_R \gets 0$
    
    \While{$|R_{\text{run}}| \neq 0$}
        \State $R_{\text{fin}} \gets \text{LLM.generate}(R_{\text{run}}, G_{\text{rollout}})$
        \State $\Delta_R \gets \Delta_R + |R_{\text{fin}}|$
        \For{$req$ in $R_{\text{fin}}$} 
            \State $R_{\text{run}}$.remove($req$)
            \State $R_{\text{comp}}$.append($req$)
        \EndFor
        \If{not $scaled\_down$}  
            % \If{$|R_{\text{comp}}| / |R| \in [0.2, 0.25, 0.3, ..., 0.5]$} \label{alg:line:scale_ratio}
            \If{$0.2\le |R_{\text{comp}}| / |R| \le 0.5$ and $\Delta_R / |R| \ge 0.05$ \label{alg:line:scale_ratio}}
                \State $\Delta_R \gets 0$
                \State $G_{\text{free}} \gets \text{PickScaleDownGPUs}(G)$
                \If{$\text{MeetScaleCriteria}(G_{\text{free}})$} \label{alg:line:check_criteria}
                    \State $G_{\text{rollout}} \gets G_{\text{rollout}} \setminus G_{\text{free}}$ \label{alg:line:scale_start}
                    \State $G_{\text{train}} \gets G_{\text{free}}$ \label{alg:line:reallocate}
                    \State $\text{MigrateRequests}(G_{\text{free}}, G_{\text{rollout}})$ \label{alg:line:migrate}
                    \State $scaled\_down \gets \text{true}$ \label{alg:line:scale_end}
                \EndIf
            \EndIf
        \EndIf
        \If{$scaled\_down$}
            \State $\text{ComputeGrad}(G_{\text{train}}, R_{\text{comp}})$ \label{alg:line:stream_train}
        \EndIf
    \EndWhile
\EndProcedure
\end{algorithmic}
\end{algorithm}

\subsection{Stream Trainer}
\label{subsec:stream-trainer}

Despite prior optimizations, long rounds still suffer from idle bubbles as
responses complete unevenly (Figure~\ref{fig:bkd_bar_sqeezer}). The \textit
{stream trainer} mitigates this with a novel stage overlap strategy that
pipelines ongoing rollouts with gradient computation,
reducing end-to-end latency while preserving the synchronous on-policy
RL semantics.

\noindent\textbf{Repurposing Rollout GPUs for Training.}
As rollout advances, GPU utilization declines. The stream trainer scales down
the number of GPUs dedicated to rollout and \emph{repurposes} the freed
devices for training. Training on repurposed GPUs proceeds with only a subset
of data-parallel replicas; gradient updates are deferred until rollout fully
completes, preserving the correctness of on-policy RL. Reference logits,
which contribute only marginally to the workload, can be computed by
temporarily swapping actor and reference model weights if needed.

Algorithm~\ref{alg:stream-trainer} outlines the general workflow of stream
trainer. It continuously monitors rollout progress and triggers GPU downscaling
once the fraction of completed requests exceeds a threshold
(Line~\ref{alg:line:scale_ratio}). When scaling criteria are met (Line~\ref
{alg:line:check_criteria}), a subset of rollout GPUs is repurposed for
training (Lines~\ref{alg:line:scale_start}-\ref{alg:line:reallocate}). To
migrate ongoing requests, \SystemName{} employs a \emph
{recomputation-based policy} (Line~\ref{alg:line:migrate}): generated tokens
are preserved while KV caches are recomputed to resume rollout on the
remaining GPUs with minimal overhead~\cite
{gao2024faststaterestorationllm,RhymeRL}. Once training instances are
launched, the stream trainer asynchronously fetches completed responses and
computes gradients in parallel with the ongoing rollout (Line~\ref
{alg:line:stream_train}).

\noindent\textbf{Scaling Criteria.} 
The trainer evaluates two criteria in Algorithm~\ref
{alg:stream-trainer} (Line~\ref
{alg:line:check_criteria}) to maximize the benefits of GPU scaling.

\textbf{\textit{1) Which GPUs to scale down?}} 
The trainer must carefully select GPUs to reassign from rollout to training,
since the two stages often rely on different communication group
topologies. To ensure correctness, tightly coupled groups, such as TP groups
used in rollout, must remain intact and cannot be split across rollout and training.
In practice, the stream trainer attempts to repurpose half of rollout GPUs 
for training. Before taking actions, it validates whether this is possible without 
splitting communication groups needed by a data-parallel replica.
If not, the scaling attempt is aborted.

\textbf{\textit{2) When to scale?}}
Downscaling rollout GPUs risks slowing response generation. By consolidating
more requests onto fewer GPUs, it enlarges per-device batch sizes,
aggravating the memory pressure for KV cache and eventually harming
rollout throughput (\S\ref{subsec:parallel-planner}). To prevent this, the
stream trainer calculates peak KV cache usage by combining historical
response length distributions with per-token cache footprints 
when the fraction of completed requests reaches
milestones between 20\% and 50\% (in 5\% increments). GPU scaling is
triggered only if the projected peak cache demand remains within memory limits
after migration. For simplicity, we omit the modest overhead introduced by
recomputation-based request migration and the minor decoding throughput
reduction after scaling (\S\ref{sub:ablation_stream_executor}).

\noindent\textbf{Overlapped Stream Execution.} 
Once the scaling criteria are met and GPUs are reassigned, the stream trainer
begins processing completed responses on the repurposed training
GPUs. Rollout and training now operate as a producer-consumer pair: rollout
generates responses, while training consumes them through a streaming model
that aligns production and consumption rates. The stream trainer
asynchronously fetches completed responses and computes gradients in parallel
with ongoing rollouts, thereby reducing the overall step time.

\noindent\textbf{Preserving On-Policy Semantics.}
A critical requirement of the stream trainer is to ensure that the gradient
computations are \emph{mathematically equivalent} to the standard on-policy
training pipeline. This guarantee is maintained in two phases. First, during
stream execution, gradients for completed responses are computed and
buffered, but \emph{no updates} are applied to model parameters or optimizer
state. We extend the underlying LLM training framework~\cite
{shoeybi2019megatron} to disable gradient synchronization during
back-propagation, ensuring strict adherence to on-policy constraints. Second,
after rollout completes, the remaining responses are distributed across all
data-parallel replicas for gradient computation and model updates. Since some
replicas may have already processed part of the workload, a naive averaging
would bias the result. To correct this, we re-normalize local gradients on
each replica by the number of samples it processed, ensuring the final update
is equivalent to that of standard on-policy training.

%% file: contents/5implementation.tex
%!TEX root=../main.tex
\section{Implementation}
\label{sec:implementation}

We implemented \SystemName{} in $\sim$6.6k lines of Python code on top of ROLL~\cite{roll}, which will be open-sourced. The system integrates
existing LLM infrastructure with lightweight extensions for rollout, reward,
and training.

\noindent\textbf{Rollout Stage.} 
\SystemName{} uses vLLM v0.8.4~\cite{vllm} as the serving backend. Each
 rollout instance supports request-level routing, allowing requests to be
 directed to specific instances. We extend vLLM's \texttt{abort\_request}  
 and \texttt{add\_request} interfaces to flexibly terminate in-progress 
 requests and resubmit them elsewhere, enabling speculative execution and migration.

\noindent\textbf{Reward Stage.} 
Reward evaluation is implemented with \texttt{ray.remote}. Code sandbox
execution and mathematical evaluation run on CPUs, with per-task timeouts of
30s and 2s, respectively. We employ \texttt{torch.cuda.Stream} to manage GPU
streams for activation computation and parameter transfers.

\noindent\textbf{Training Stage.}
Actor training is built on Megatron-LM v0.12.2~\cite
{shoeybi2019megatron}, with optimizer states partitioned across GPUs. In the
stream trainer, gradients are computed without loading optimizer states.
Gradient tensors are offloaded to host and later reloaded into GPU memory
when synchronizing across all GPUs for final gradient computation and updates.

%% file: contents/6evaluation.tex
%!TEX root=../main.tex
\section{Evaluation}
\label{sec:eval}

\begin{figure*}[tb]
    \centering
    \begin{subfigure}{0.32\textwidth}
        \centering
        \includegraphics[width=\linewidth]{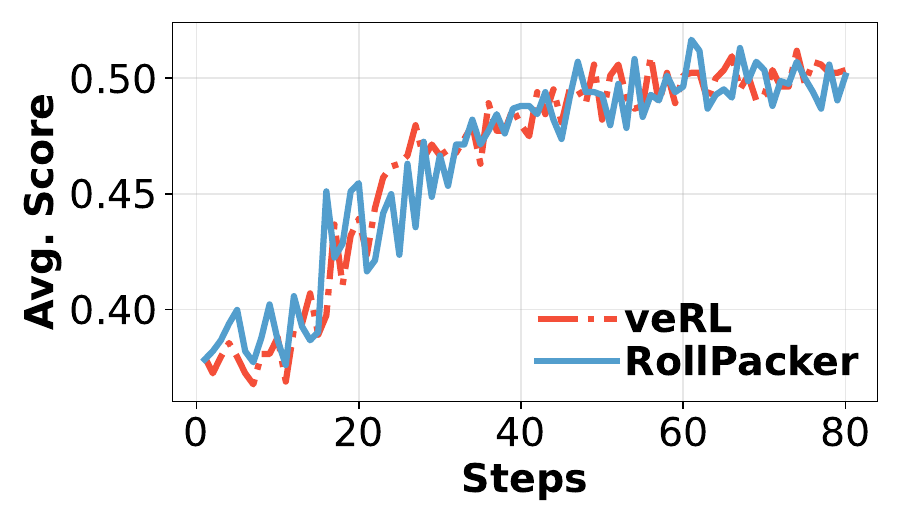}
        \caption{Qwen2.5-7B}
        \label{subfig:e2e_score_7b}
    \end{subfigure}
    \begin{subfigure}{0.32\textwidth}
        \centering
        \includegraphics[width=\linewidth]{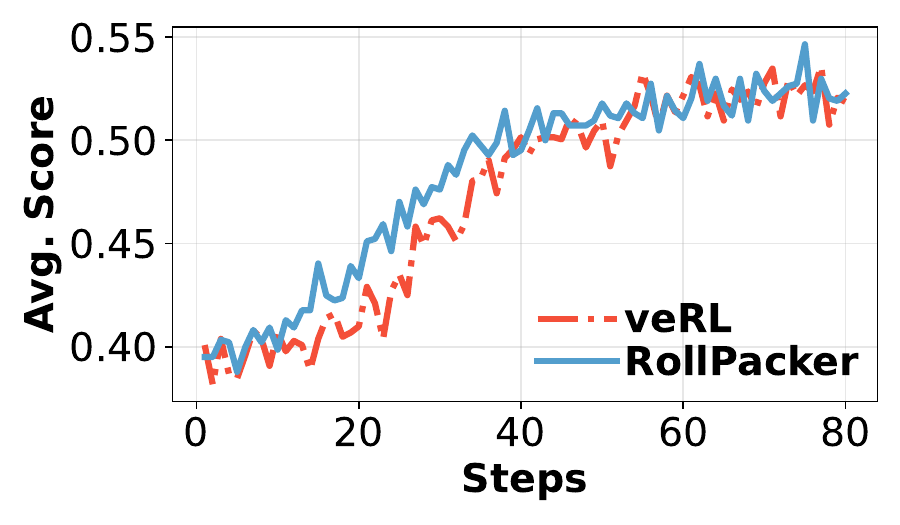}
        \caption{Qwen2.5-14B}
        \label{subfig:e2e_score_14b}
    \end{subfigure}
    \begin{subfigure}{0.32\textwidth}
        \centering
        \includegraphics[width=\linewidth]{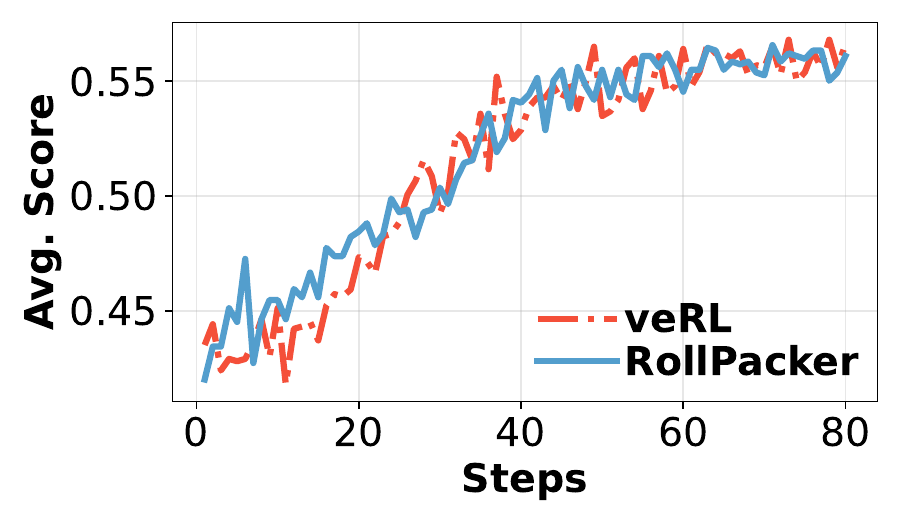}
        \caption{Qwen2.5-32B}
        \label{subfig:e2e_score_32b}
    \end{subfigure}
    \caption{
        The average validation score for training Qwen2.5-7B, 14B and 32B model with veRL and \SystemName{}.
    }
    \vspace{-10pt}
    \label{fig:e2e_score}
\end{figure*}

\begin{figure*}[tb]
    \centering
    \begin{subfigure}{0.32\textwidth}
        \centering
        \includegraphics[width=\linewidth]{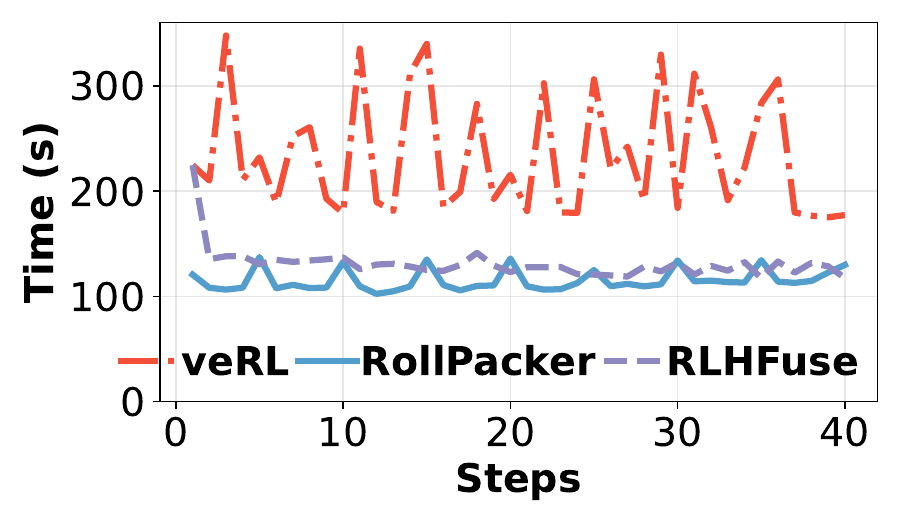}
        \caption{Qwen2.5-7B}
        \label{subfig:step_total_7b}
    \end{subfigure}
    \begin{subfigure}{0.32\textwidth}
        \centering
        \includegraphics[width=\linewidth]{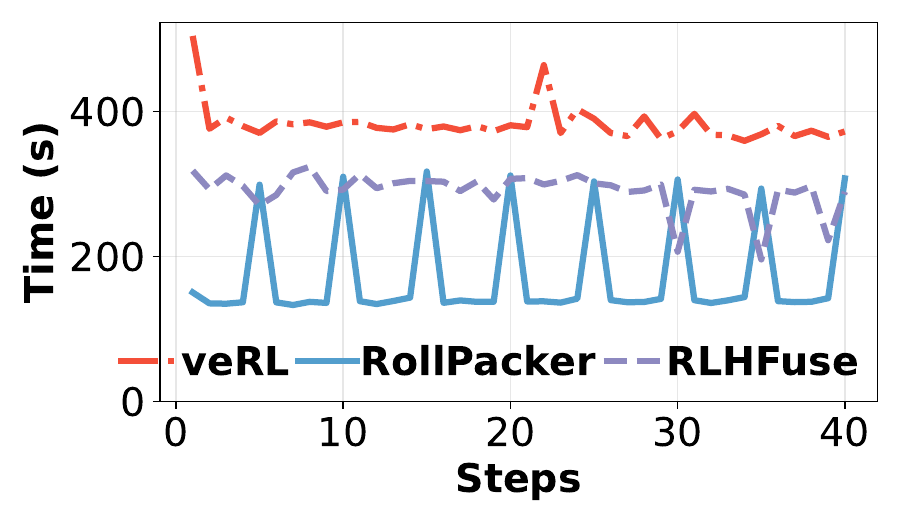}
        \caption{Qwen2.5-14B}
        \label{subfig:step_total_14b}
    \end{subfigure}
    \begin{subfigure}{0.32\textwidth}
        \centering
        \includegraphics[width=\linewidth]{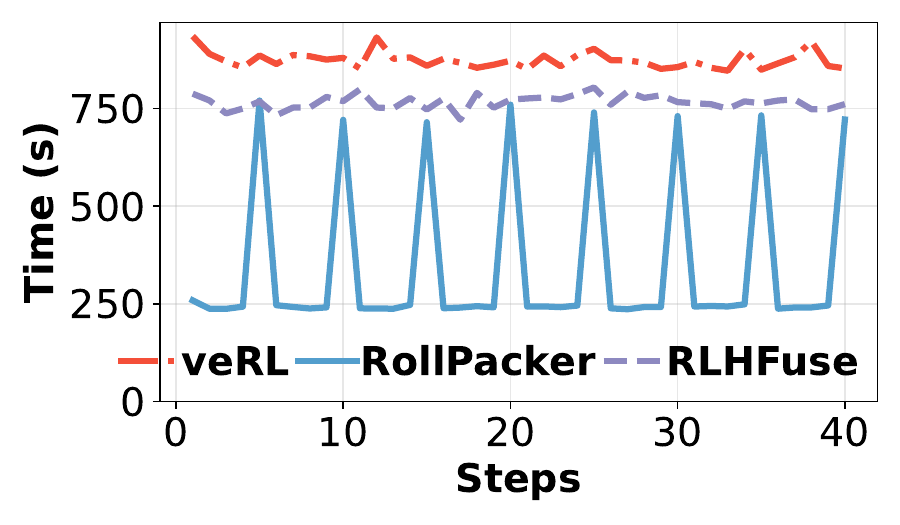}
        \caption{Qwen2.5-32B}
        \label{subfig:step_total_32b}
    \end{subfigure}
    \caption{
        The step time for training Qwen2.5-7B, 14B and 32B model with veRL, RLHFuse and \SystemName{}.
    }
    \label{fig:step_total}
    \vspace{-10pt}
\end{figure*}

\begin{figure*}[tb]
    \centering
    \begin{subfigure}{0.32\textwidth}
        \centering
        \includegraphics[width=\linewidth]{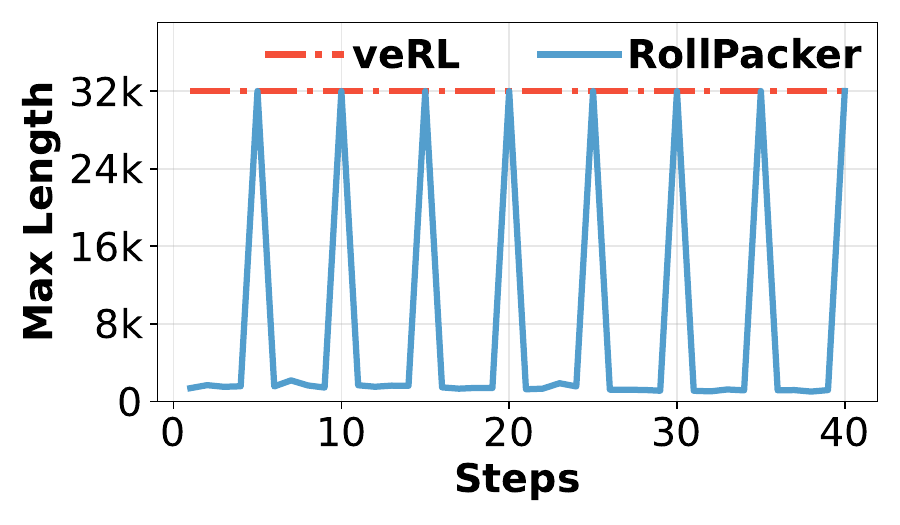}
        \caption{Max Response Length [Qwen2.5-32B]}
        \label{subfig:max_length_32b}
    \end{subfigure}
    \begin{subfigure}{0.32\textwidth}
        \centering
        \includegraphics[width=\linewidth]{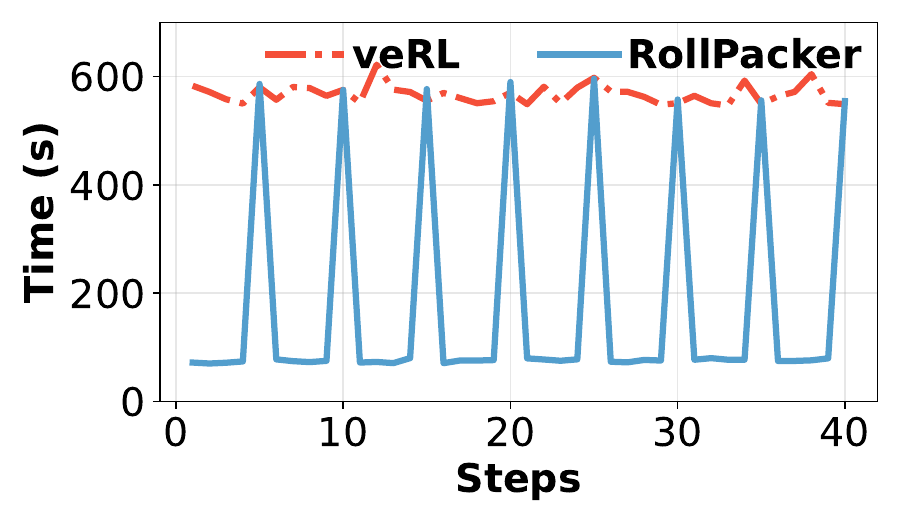}
        \caption{Rollout Time [Qwen2.5-32B]}
        \label{subfig:step_gen_32b}
    \end{subfigure}
    \begin{subfigure}{0.32\textwidth}
        \centering
        \includegraphics[width=\linewidth]{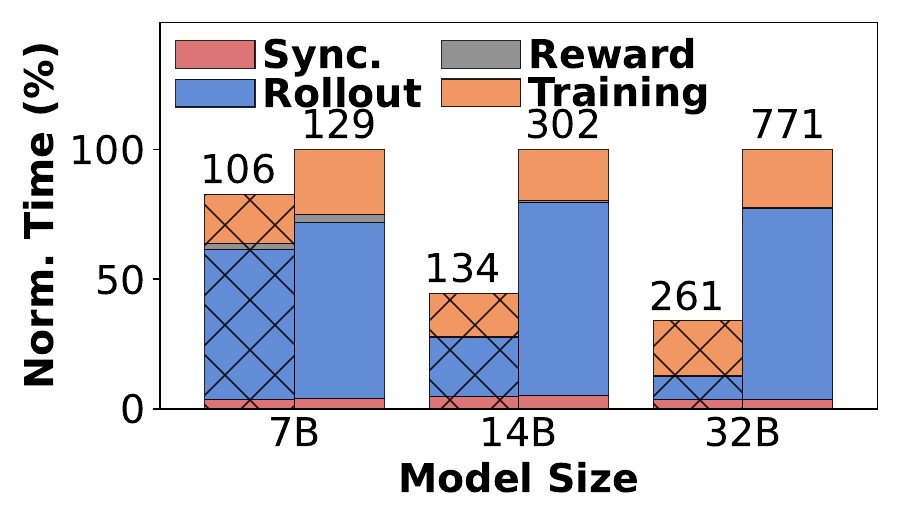}
        \caption{Step Time Breakdown}
        \label{subfig:step_bkd}
    \end{subfigure}
\caption{Breakdown of training step time in \SystemName{}. 
(a) Maximum response length per training step for Qwen2.5-32B/32k. 
(b) Total rollout time per training step for Qwen2.5-32B/32k. 
(c) Breakdown of step time for different models, comparing short rounds (hatched bars) with long rounds (solid bars). The time for each component is normalized to the total time of the long round for that model. Absolute step times are displayed as labels on each bar.}
    \label{fig:performance_bkd}
\end{figure*}

We evaluate \SystemName{} on Qwen2.5 models with 7B-32B parameters using a diverse benchmark of real-world datasets. In \S\ref{subsec:e2e-eval}, we
compare \SystemName{} against existing RL post-training systems in terms of
validation accuracy and training time. We then break down performance across
pipeline stages in \S\ref{sub:performance-breakdown} and microbenchmark the
three optimization designs in \S\ref{sub:impact-of-key-parameters}-\S\ref
{sub:ablation_stream_executor}. \S\ref{sub:large_scale_result} presents a scalability analysis.

\noindent\textbf{Cluster Setup.} 
We deploy \SystemName{} on an H800 cluster with 16 nodes (128 GPUs total), connected via 
400 Gbps InfiniBand.

\noindent\textbf{Models.}  
We use the Qwen2.5~\cite{qwen} family with 7B, 14B, and 32B
parameters, configured with maximum response lengths of 8k, 16k, and 32k
tokens, respectively. End-to-end evaluation is conducted in a multi-task 
setting with a uniform mix of datasets spanning mathematics~\cite
{he2025deepmath}, code generation~\cite{xu2025kodcode}, and multi-subject
question answering, using rule-based, code sandbox, and
LLM-as-a-Judge reward workers, respectively.

\noindent\textbf{Training Configurations.} 
Unless otherwise noted, we adopt synchronous RL training with $P_0=128$
and $R_0=8$. Actor and reference models are of the same size, and
Qwen2.5-7B-Instruct is used as the judge. 
Parallelism strategies and resource
allocations vary with model size. The 7B, 14B, and 32B models are
trained on 16, 32, and 64 GPUs. Their rollout TP is set to 1, 2, and 2, while
training configurations (TP, PP, CP) are (2,1,1), (2,2,2), and
(4,1,4), respectively.

\noindent\textbf{Metrics.}
We report validation accuracy across training steps and measure end-to-end
training time to evaluate both effectiveness and efficiency.

\subsection{End-to-End Evaluation}
\label{subsec:e2e-eval}
We compare the end-to-end performance of \SystemName{} with state-of-the-art synchronous RL post-training systems.
\begin{itemize}[leftmargin=*]
    \setlength\itemsep{-0.2em}
    \item \textbf{veRL}~\cite{verl} proposes a hybrid programming model for the RL post-training pipeline and provides a optimized 3D-HybridEngine to improve the rollout and training efficiency. 
    \item \textbf{RLHFuse}~\cite{rlhfuse} pipelines the reward and reference model inference with the rollout stage. We strength RLHFuse with \textit{stream trainer} and \textit{asynchronous reward computation}. 
\end{itemize}

    % It is a synchronous RL post-training system tackling the long-tail rollout.

\noindent\textbf{Validation Performance.}
Figure~\ref{fig:e2e_score} presents the average validation scores of \SystemName{} and veRL, demonstrating that tail batching does not compromise training accuracy across different model sizes and response lengths. Moreover, \SystemName{} achieves faster convergence at early stage, and we hypothesize it is due to the more balanced response length distribution.

\noindent\textbf{End-to-End Latency.} Figure~\ref{fig:step_total} reports the training step time of each model in the first 40 steps for a clear illustration. Overall, \SystemName{} outperforms veRL and RLHFuse across all three LLMs. Compared with veRL, \SystemName{} achieves speedups of $2.03\times$, $2.22\times$, and $2.56\times$ for three LLMs, respectively. Against RLHFuse, the speedups are $1.14\times$, $1.68\times$, and $2.24\times$. Owing to the reward scheduler and stream trainer, both \SystemName{} and RLHFuse maintain advantages over veRL in long rounds. However, the overlapping benefits diminish as response length increases (see Figure~\ref{subfig:step_total_32b}), since the proportion of rollout time grows in long rounds. In short rounds, \SystemName{} achieves $2.1\times$-$3.6\times$ speedup over veRL and $1.2\times$-$3.2\times$ speedup over RLHFuse, owing to tail batching.

\begin{table}[tb]
    \centering
    \caption{End-to-end training speedup breakdown.}
    % \scriptsize
    \resizebox{0.95\linewidth}{!}{
    % \begin{tabularx}{0.\linewidth}{lccc}
    \begin{tabularx}{\linewidth}{l *{3}{>{\centering\arraybackslash}X}}
    \toprule
        \textbf{Method} & \textbf{Qwen2.5-7B/8k} & \textbf{Qwen2.5-14B/16k} & \textbf{Qwen2.5-32B/32k}\\ 
        \midrule
        veRL Baseline & 1.00~~~~ & 1.00~~~~ & 1.00~~~~ \\ 
        + Tail Batching & 1.30$\times$  & 1.48$\times$ & 2.21$\times$ \\ 
        + Reward & 2.01$\times$ & 1.99$\times$ & 2.48$\times$ \\ 
        + Parallelism & 2.01$\times$ & 2.02$\times$ & 2.52$\times$ \\ 
        + Trainer & 2.03$\times$ & 2.22$\times$ & 2.56$\times$  \\ 
        \bottomrule
    \end{tabularx}}
    \label{tab:ablation_e2e}
\end{table}

\subsection{Performance Breakdown}
\label{sub:performance-breakdown}
\noindent\textbf{Improvement Breakdown.}
Table~\ref{tab:ablation_e2e} presents a detailed breakdown of the cumulative speedup from our proposed techniques across different model sizes and response lengths. 

\begin{itemize}[leftmargin=*]
    \setlength\itemsep{-0.2em}
    \item The tail batching effectively reduces rollout overhead, and its benefits become more pronounced as response length increases. In particular, \SystemName{} achieves up to $2.21\times$ speedup for Qwen2.5-32B/32k setting.
    \item The reward scheduler is particularly beneficial for short rollouts, providing a $71\%$ performance uplift (from $1.30\times$ to $2.01\times$) with an 8k response length. It also retains its advantages for longer responses, contributing a $27\%$ improvement (from $2.21\times$ to $2.48\times$) at a 32k response length.
    \item The parallelism planner is effective under high memory pressure, a scenario typical for large models and long sequences. In the Qwen2.5-32B/32k setting, it provides an additional $4\%$ speedup (from $2.48\times$ to $2.52\times$).
    \item The stream trainer reduces step time by overlapping rollout and gradient computation. It delivers a substantial $20\%$ performance improvement (from $2.02\times$ to $2.22\times$) for the Qwen2.5-14B/16k, where rollout and training time are well balanced for pipelining. For other settings, the stream trainer consistently yields positive speedup gains.
\end{itemize}

\noindent\textbf{Training Step Breakdown.}
Figure~\ref{fig:performance_bkd} breaks down the training time to analyze the performance of \SystemName{}. Figures~\ref{subfig:max_length_32b} and~\ref{subfig:step_gen_32b} compare \SystemName{} against veRL on maximum response length and rollout time. While rollout times are comparable in long rounds, \SystemName{} demonstrates a significant advantage in short rounds. It substantially reduces the maximum response length (Figure~\ref{subfig:max_length_32b}), leading to up to a $7.8\times$ speedup in average rollout time (Figure~\ref{subfig:step_gen_32b}). The aggregated step time breakdown is shown in Figure~\ref{subfig:step_bkd}, with hatched bars for short rounds and solid bars for long rounds. In long rounds, the rollout stage progressively dominates the total step time. In contrast, the time savings from the shorter rollouts become more significant, effectively lowering the average step time across all rounds. Next, we investigate each system component and quantify its individual contribution under various conditions. The results of these microbenchmarks are presented from \S\ref{sub:impact-of-key-parameters} to \S\ref{sub:ablation_stream_executor}.

\subsection{Sensitivity Analysis of Tail Batching}
\label{sub:impact-of-key-parameters}

\begin{figure}[tb]
\centering
\includegraphics[width=\linewidth]{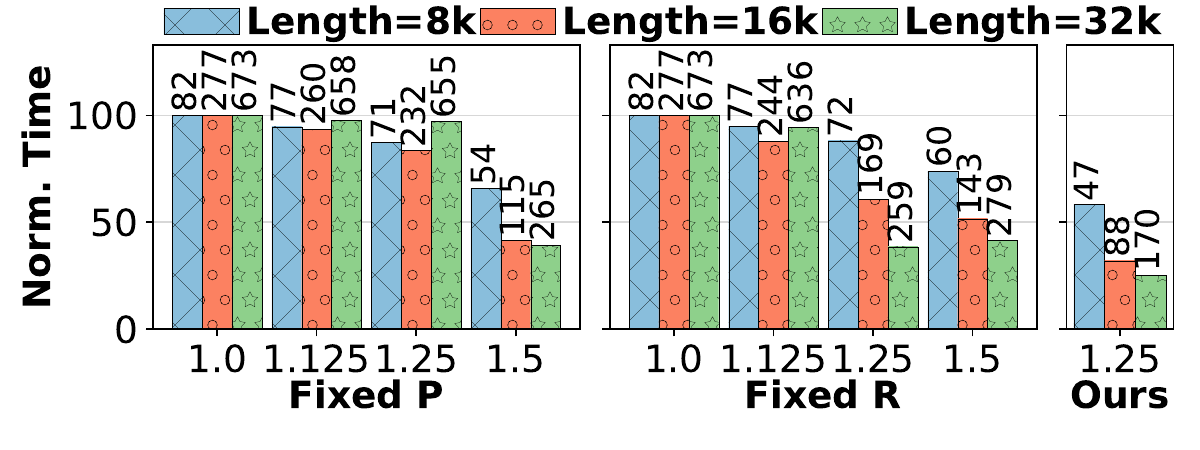}
\caption{
\textbf{[Tail Batching]} The rollout time of different configurations.
We set $P$ and $R$ fixed to $P_0$, and $R_0$, respectively, while changing $\eta$ for the another parameter.
Each bar represents the average iteration time of a full period of consecutive short and long rounds, normalized to the baseline without tail batching, and is annotated with the actual time.
}
\label{fig:ablation_squeezer}
\end{figure}

Figure~\ref{fig:ablation_squeezer} shows the rollout time of different configurations of tail batching. We first fix the number of prompts to $P_0$ and set different $\eta$ for the number of responses per prompt $R$.
We then fix the number of responses per prompt and discover the impact of $R$.
We compare them with our chosen configuration of $\eta=1.25$ in \SystemName{}.

\noindent\textbf{Impact of $R$.}
With a fixed $P = P_0$, we increase $\eta$ for $R$ from $1.0$ to $1.5$. As the number of responses per prompt increases, we can discard long-tail responses to reduce rollout time. However, some difficult prompts consistently produce long responses.  yield long responses. A substantial reduction in rollout time is observed only when $\eta$ is increased to $1.5$.

\noindent\textbf{Impact of $P$.}
With a fixed $R = R_0$, we increase the number of prompts $P$. We collect the first first $P_0$ prompts and drops the remaining prompts with long responses to the long-prompt queue. When we increase $\eta$ for $P$, the frequency of long rounds increases accordingly, which negates the benefits of time reduction from short rounds. For example, with a response length of 32k, we observe that the average rollout time increases when $\eta$ is raised from $1.25$ to $1.5$.

Based on the above analyses of $P$ and $R$, we fix $\eta = 1.25$ for both. Under this setting, tail batching improves the average rollout speed by up to $3.9\times$ and outperforms the fixed-$P_0$ and fixed-$R_0$ settings by up to $1.5\times$ and $1.6\times$, respectively.

\begin{figure*}[tb]
    \centering
    \begin{subfigure}{0.38\textwidth}
        \centering
        \includegraphics[width=0.98\linewidth]{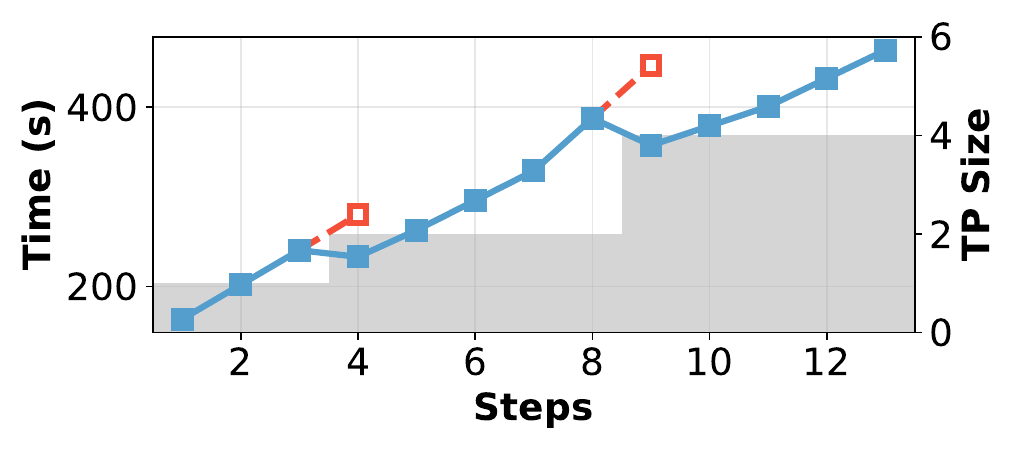}
        \caption{Dynamic TP}
        \label{fig:ablation_adpt_tp}
    \end{subfigure}
    \begin{subfigure}{0.27\textwidth}
        \centering
        \includegraphics[width=\linewidth]{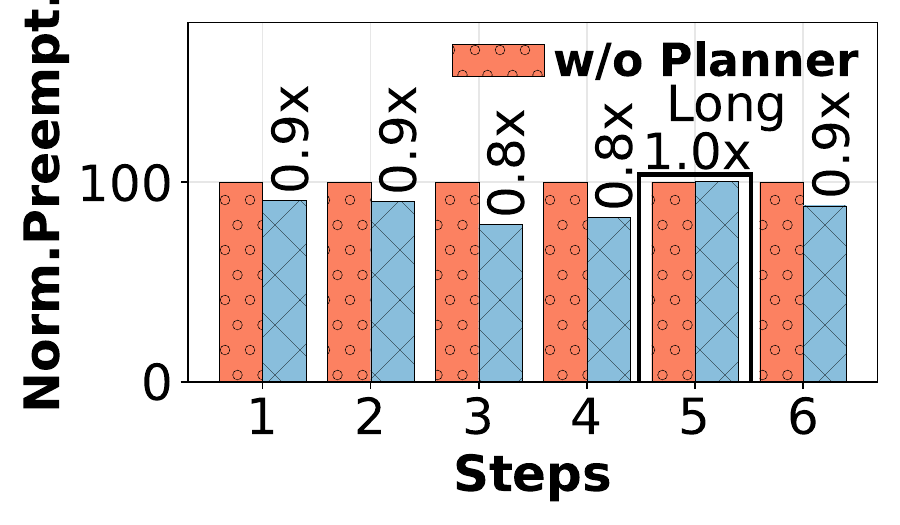}
        \caption{Preemption Count}
        \label{subfig:ablation_preempts}
    \end{subfigure}
    \begin{subfigure}{0.27\textwidth}
        \centering
        \includegraphics[width=\linewidth]{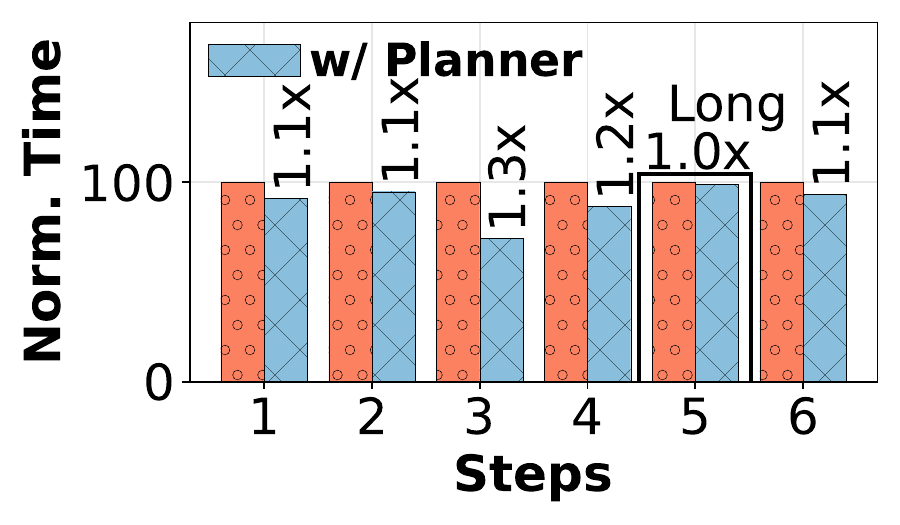}
        \caption{Rollout Time}
        \label{subfig:ablation_times}
    \end{subfigure}
    \caption{\textbf{[Parallelism Planner]} (a) Rollout time (line, left axis) and corresponding TP size (bar, right axis) when training Qwen2.5-14B. The response length increases linearly from 8k to 32k in 2k increments. Red dashed lines indicate rollout time without TP adjustment. (b)–(c) Normalized preemption count and rollout time per step with and without the parallelism planner. The fifth step corresponds to the long round.
    }
    \label{fig:ablation_para_planner}
\end{figure*}

\subsection{Parallelism Planner}
\label{sub:impact-parallel-planner}

\noindent\textbf{Dynamic TPs in LLM Generation.}
The parallelism planner adaptively adjusts the TP size for LLM generation in the rollout stage based on the response length distribution. To analyze this behavior, we measure the rollout time of training Qwen2.5-14B on 16 GPUs across training steps. Initially, we fix the TP size to 1, then gradually increase the response length from 8k to 32k by 1k per training step. We present the average rollout time (left) and the optimal TP size (right) in Figure~\ref{fig:ablation_adpt_tp}. Specifically, at steps 3 and 8, the parallelism planner increases the TP size to 2 and 4, respectively, to reduce rollout time. We also compare the iteration time when changing the TP size (blue solod line) versus keeping it fixed (red dashed line), and observe a clear reduction in rollout time due to adaptive TP selection. Overall, the parallelism planner achieves an average $1.9\times$ speedup compared to a baseline with a fixed TP size as 1.

\noindent\textbf{Preemptions and Rollout Latency.}
Figures~\ref{subfig:ablation_preempts}-\ref{subfig:ablation_times} show the number of preemptions and the rollout time per step, with and without the parallelism planner, when the maximum response length is fixed at 32k and the initial TP size is set to 2. Both the preemption count and rollout time are normalized to the values obtained without the parallelism planner. As shown in Figure~\ref{subfig:ablation_preempts}, the parallelism planner reduces the preemption count in short rounds by an average of $13.8\%$. Figure~\ref{subfig:ablation_times} shows that the parallelism planner can speedup the rollout time in short rounds by $1.11\times$-$1.28\times$.

\begin{figure*}[tb]
    \centering
    \begin{subfigure}{0.28\textwidth}
        \centering
        \includegraphics[width=\linewidth]{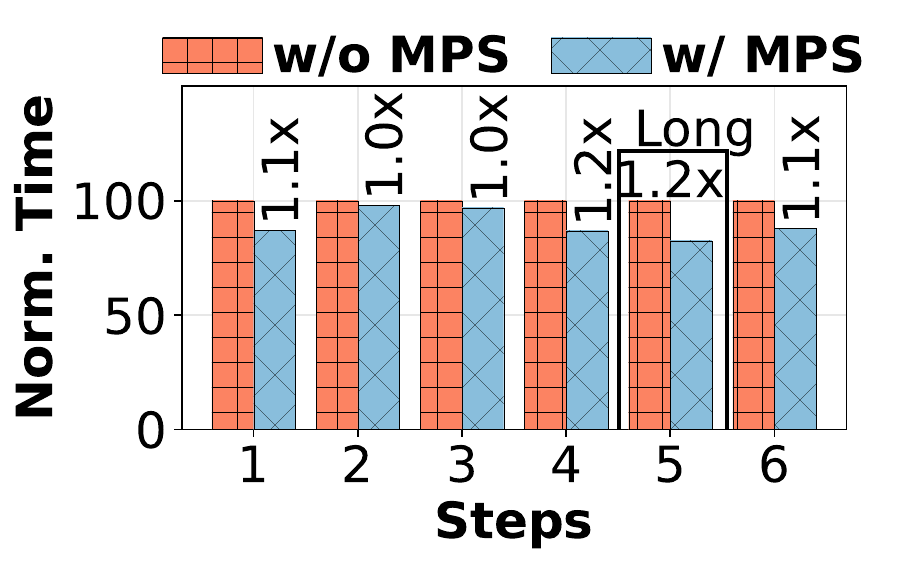}
        \caption{MPS}
        \label{subfig:mps_time}
    \end{subfigure}
    \begin{subfigure}{0.28\textwidth}
        \centering
        \includegraphics[width=\linewidth]{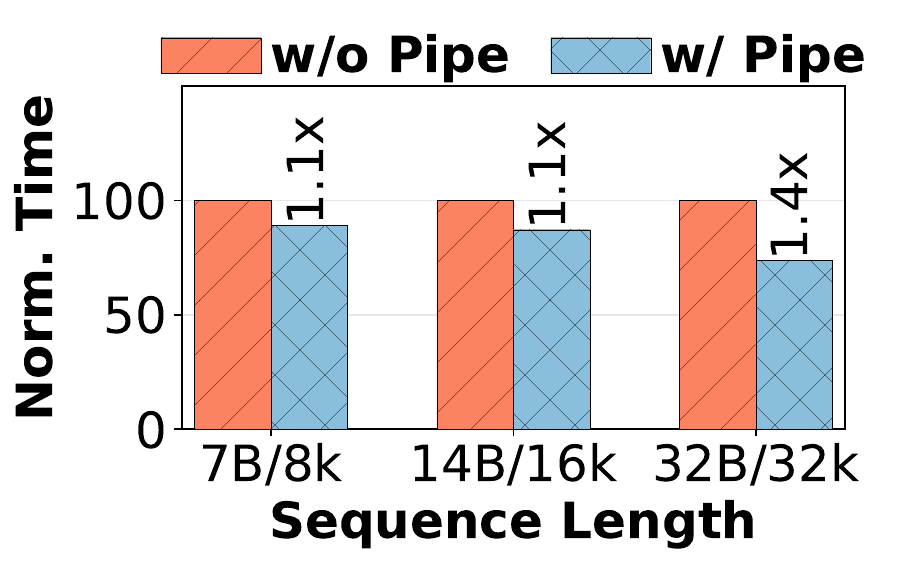}
        \caption{Layer-wise Pipeline}
        \label{subfig:pipe_ludge}
    \end{subfigure}
    \begin{subfigure}{0.38\textwidth}
        \centering
        \includegraphics[width=.95\linewidth]{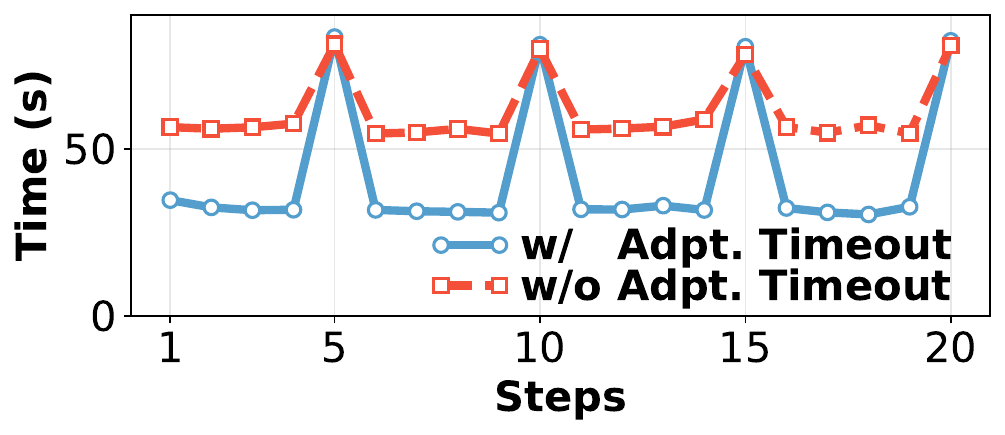}
        \caption{Adaptive Timeout}
        \label{fig:ablation_code_adaptive_time_limit}
    \end{subfigure}
    \caption{\textbf{[Reward Scheduler]}
        \textit{LLM-as-a-Judge}: 
        (a) Normalized time for each step w/ and w/o \texttt{MPS}. 
        The fifth step is the long round. 
        (b) Normalized time for reward computation with different sequence lengths w/ and w/o pipelined execution. \textit{Code sandbox}: (c) The combined time of rollout and asynchronous reward computation w/ and w/o adaptive timeout. 
    }
    \label{fig:ablation_rw_sched}
\end{figure*}

\subsection{Reward Scheduler}
Compared with synchronous reward computation, asynchronous reward computation yields speedups of $1.48\times$, $1.35\times$, and $1.18\times$ on three LLMs, respectively. 
We analyze the reward scheduler for LLM-as-a-Judge and code tasks. 
To isolate and evaluate its effectiveness, each subsequent experiment is conducted on a specific task rather than a mixture.

\label{sub:impact-reward-scheduler}
\noindent\textbf{GPU Sharing in Judge LLM.} 
In the reward scheduler, we colocate the judge LLM and actor LLM on the same GPU to improve the resource utilization. However, concurrent executions on a single GPU can suffer interference due to contention for shared computational resources, resulting in degraded performance. To mitigate this, we leverage \texttt{MPS}. Figure~\ref{subfig:mps_time} reports the step time with and without \texttt{MPS} when training Qwen2.5-7B/8k. We observe that \texttt{MPS} consistently reduces step time, yielding up to a $1.25\times$ speedup. These results demonstrate the effectiveness of \texttt{MPS} in GPU sharing.

% layerbylayer pipeline

\noindent\textbf{Pipelined Judge LLM Execution.}
We employ a layer-wise pipelined execution scheme that offloads the weights of the judge LLM to CPU memory and overlaps weight transmission with GPU activation computation.
% to mitigate this issue. 
When colocating the judge LLM and actor LLM, we reserve GPU memory for the judge LLM, which requires offloading at least half of its weights to the CPU in order to perform reward computation for responses of maximum length. Although the layer-wise scheme enables execution of the judge LLM under memory constraints, it incurs substantial weight transmission overhead. Figure~\ref{subfig:pipe_ludge} compares pipelined and non-pipelined execution in terms of reward computation overhead across different sequence lengths. Pipelined execution yields up to a $1.4\times$ speedup when the maximum response length reaches 32k tokens, as the larger activations demand more GPU memory and necessitate offloading more LLM weights. Overall, these results demonstrate that pipelining simultaneously reduces memory consumption and reduces reward computation time.

\noindent\textbf{Adaptive Timeout for Code.} we employ an adaptive timeout to alleviate the code sandbox execution overhead. We compare reward computation overhead under adaptive and fixed timeout with Qwen2.5-7B/8k. Figure~\ref{fig:ablation_code_adaptive_time_limit} shows the combined duration of rollout and asynchronous reward computation at each step. The adaptive timeout substantially reduces the unnecessary timeouts
% and reward computation overhead  
in the short rounds. In the long round, the reward computation can be overlapped with long-tail rollouts. These results highlight the effectiveness of the adaptive timeout in improving reward computation efficiency, achieving an average speedup of 1.6\(\times\) across all steps.

\subsection{Performance Analysis of Stream Trainer}
\label{sub:ablation_stream_executor}
We evaluate the effectiveness of the stream trainer using Qwen2.5-7B/8k on mathematical datasets for simplicity.
% to clearly analyze its benefits.

\noindent\textbf{Adaptive Criteria for GPU Scaling.} 
The stream trainer monitors the response length distribution and decides when to perform GPU scaling adaptively. 
% A straightforward approach is to apply a fixed criterion to trigger GPU scaling.  
To highlight the advantages of this adaptive criterion, we use a baseline without GPU scaling and fixed-trigger baselines where scaling is applied when the number of completed prompts reaches 20\%, 30\%, or 40\%.  
Since short rounds involve less frequent GPU scaling, we report the average speedup over five long rounds. Asynchronous fetching is enabled for all GPU scaling baselines.  
We observe that even with fixed criteria, GPU scaling reduces end‑to‑end training time and the migration overhead does not exceed 3 seconds, and GPU scaling decreases the decoding throughput ranges within 1\%. Furthermore, adaptive GPU scaling outperforms all fixed‑criteria baselines, achieving a $1.08\times$ speedup over the baseline without GPU scaling as depicted in Table~\ref{tab:ablation_scale_ratio}.

\noindent\textbf{Asynchronous Fetching.} 
The stream trainer asynchronously fetches completed prompts from the rollout stage to compute gradients.  
To highlight the benefits of this streaming behavior, we compare it against baseline approaches that fetch all available completed prompts only once, with the number of fetched samples capped at 20\%, 30\%, 40\%, or 50\% of the total.  
Table~\ref{tab:ablation_stream_fetching} reports the end-to-end training step time for different fixed fetch ratios versus asynchronous prefetching. Compared to fixed-size fetching, the stream trainer achieves up to a 14\% reduction in end-to-end step time.

\begin{table}[tb]
\centering
\caption{\textbf{[Stream Trainer]} The impact of GPU scaling.}
\resizebox{0.95\linewidth}{!}{
\begin{tabularx}{\linewidth}{l *{5}{>{\centering\arraybackslash}X}}
\toprule
\textbf{} & \textbf{w/o} & \textbf{20\%} & \textbf{30\%} & \textbf{40\%} & \textbf{Adpt.} \\
\midrule
\multirow{2}{*}{Step Time (s)} & 124.2 & 122.7 & 118.2 & 119.8 & 115.2 \\
 & 1.00$\times$ & 1.01$\times$ & 1.05$\times$ & 1.04$\times$ & 1.08$\times$ \\
\bottomrule
\end{tabularx}}
\label{tab:ablation_scale_ratio}
\end{table}

\begin{table}[tb]
\centering
\caption{\textbf{[Stream Trainer]} The impact of async fetching.}
\resizebox{0.95\linewidth}{!}{
\begin{tabularx}{\linewidth}{l *{5}{>{\centering\arraybackslash}X}}
\toprule
\textbf{} & \textbf{Stream} & \textbf{20\%} & \textbf{30\%} & \textbf{40\%} & \textbf{50\%} \\
\midrule
\multirow{2}{*}{Step Time (s)} & 115.2 & 128.6 & 129.5 & 133.8 & 132.0 \\
 & 1.00$\times$ & 0.90$\times$ & 0.89$\times$ & 0.86$\times$ & 0.87$\times$ \\
\bottomrule
\end{tabularx}}
\label{tab:ablation_stream_fetching}
\end{table}

\subsection{Performance Scalability}
\label{sub:large_scale_result}
We conduct a scalability analysis for the Qwen2.5-14B/16k. 
We scale the batch size from 128 to 512 along with the corresponding computational resources. Throughput is measured following~\cite{rlhfuse}, defined as the average number of samples processed per second, and is averaged over 20 consecutive training steps. Figure~\ref{fig:sacling_throughput} shows that \SystemName{} maintains strong performance at large scale, utilizing up to 128 GPUs. Compared with veRL, \SystemName{} consistently achieves a $2.2\times$ throughput increase. When scaling up resources by $2\times$, \SystemName{} delivers $\sim$$1.5\times$ throughput improvement, with the smaller gain attributed to the increased training time for larger batch sizes.

%% file: contents/7relatedworks.tex
\section{Discussion}
\label{sec:discussion}
Here, we discuss how \SystemName{}'s design benefits to other policy optimization algorithms and off-policy RL algorithms. We also discuss the potential limitations of \SystemName{}. 

\noindent\textbf{Extend to Other Policy Optimization Algorithms.} Many algorithmic studies~\cite{DAPO,GRESO,MoPPS} have extended GRPO to improve sample efficiency. A representative variant is DAPO, which performs oversampling and discards prompts with zero reward variance. Similar to tail batching, DAPO launches more prompts than the batch size during the rollout stage. To integrate tail batching with DAPO, we set a maximum number of active requests for each LLM instance and continuously issue new requests. The termination criteria follow DAPO specifications. Prompts with zero reward variance are excluded from the long-prompt queue, while other unfinished prompts are retained in the queue for subsequent processing.

\noindent\textbf{Extend to Asynchronous Systems.} In asynchronous off-policy post-training, synchronous RL training is not required to optimize the samples in the long-prompt queue.  
We can instead leverage existing off-policy algorithms, including one-off pipeline~\cite{deepscaler2025}, partial rollouts~\cite{kimiteam2025kimik15}, and fully asynchronous training~\cite{areal}, to process the prompts in the queue.  
For example, rollouts for unfinished prompts can simply be continued in the next training step.  
In addition, the reward scheduler and stream trainer can reduce the overhead of reward computation and LLM training for asynchronous training, respectively.

\noindent\textbf{Potential Limitations.} We utilize \texttt{MPS}~\cite{nvidia_mps} to enable spatial GPU sharing, which does not guarantee error isolation. As future work, we plan to explore \texttt{Green Contexts}~\cite{nvidia_mps_docs} to improve fault tolerance. The parallelism planner currently focuses only on TP and does not optimize for expert parallelism. Once expert parallelism is enabled, the optimization space can be further expanded.

\section{Related Works} 
\label{sec:related}
\noindent\textbf{RL Post-training Frameworks.} Many frameworks have been proposed to accelerate RL post-training. Early efforts~\cite{openrlhf,deepspeedchat,Harper_NeMo_a_toolkit,Harper_NeMo_a_toolkit,lei2024puzzle} aim to orchestrate the complex workflow of RL post-training. Later, veRL~\cite{verl} introduces a hybrid-controller design to improve resource utilization, while DistFlow~\cite{distflow} adoptes a multi-controller approach to enhance scalability. RLHFuse~\cite{rlhfuse} fuses the generation and inference stages to reduce training time, and Realhf~\cite{realhf} optimizes parallelism strategies to improve system throughput. To mitigate long-tail rollouts, AReal~\cite{areal}, StreamRL~\cite{StreamRL}, and RhymeRL~\cite{RhymeRL} introduce asynchronous RL post-training with tailored system optimizations to increase throughput. 
Unlike existing RL frameworks, \SystemName{} can alleviate long-tail rollouts even in synchronous RL training.

\noindent\textbf{LLM Training and Inference.} LLM training necessitates a range of parallelism strategies. Data parallelism (DP)~\cite{paszke2019pytorch, alex2018horovod, rajbhandari2020zero, zhao2023PyTorchFSDP} replicates model weights across GPUs and partitions training samples among model replicas. Tensor parallelism (TP)~\cite{shoeybi2019megatron, xu2023efficient, wang2022tesseract, bian2021maximizing} partitions computation within a model layer, while pipeline parallelism (PP)~\cite{huang2019gpipe, narayanan2019pipedream} partitions the model across layers. Context parallelism (CP)~\cite{li2022SequenceParallelism, korthikanti2022ReducingActivation, MegatronCP, liu2023ring} partitions input sequences and requires dedicated attention-layer optimizations. In LLM inference, TP and DP are typically used to reduce latency overhead. Many inference optimization techniques focus on accelerating attention computation~\cite{flashattention,vllm} and optimizing KV cache management~\cite{sglang,vllm,yu2022orca}. These strategies reduce memory consumption from different dimensions and achieve significant throughput improvements.

\noindent\textbf{Rollout Optimization.} 
Many recent works aim to optimize the rollout stage of RL post-training to improve training efficiency. DAPO~\cite{DAPO} proposes a dynamic sampling technique to filters out prompts with zero reward variance and terminates the rollout stage after collected enough responses. SPEED-RL~\cite{SPEEDRL} estimates the difficulty of each prompt, then selects those with desirable pass rates for further response generation. GRESO~\cite{GRESO} leverages reward dynamics to remove zero-variance prompts before rollout, while MoPPS~\cite{MoPPS} models prompt success rates to predict prompt difficulty.
These techniques expedite the model convergence in RL post-training by prioritizing high-quality prompts. \SystemName{} aims to improve the rollout speed, thus reducing end-to-end training latency.

\begin{figure}[tb]
\centering
\includegraphics[width=0.8\linewidth]{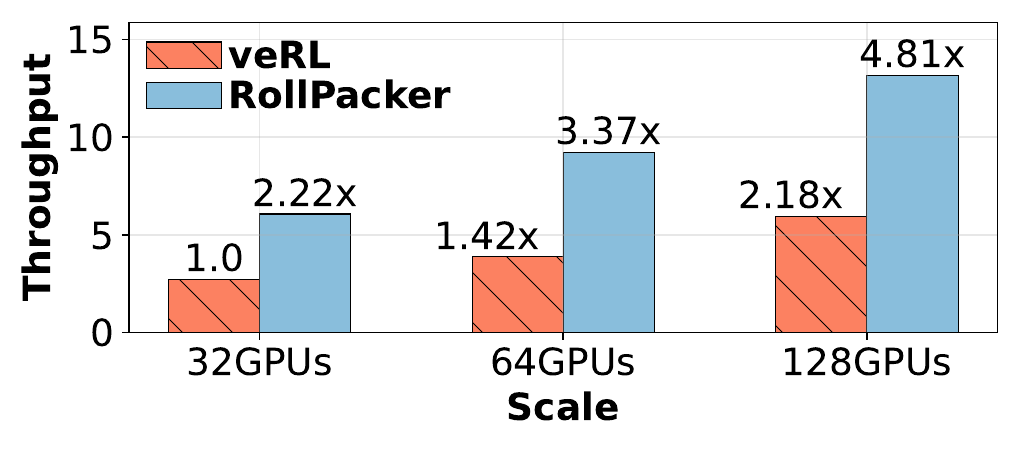}
\vspace{-2pt}
\caption{End-to-end throughput (samples/second) of training a Qwen2.5-14B model at scale.}
\label{fig:sacling_throughput}
\end{figure}

%% file: contents/8conclustion.tex
\section{Conclusion}

This paper presents \SystemName{}, a novel RL post-training system designed to expedite synchronous RL training. We propose \textit{tail batching} to alleviate long-tail rollouts and enhance resource utilization. In conjunction with the tail batching, we design parallelism planner, reward scheduler, and stream trainer that optimize the rollout, reward, and training stages respectively. Extensive experiments demonstrate the effectiveness of \SystemName{} in training efficiency against baselines. 
% compared to existing approaches.